\documentclass[%
reprint, superscriptaddress,
showpacs,
 amsmath,amssymb,
 aps,
]{revtex4-1}
\usepackage[svgnames]{xcolor}
\usepackage{epsfig,float,subfigure,mathrsfs,mathtools,ulem}
\definecolor{darkbrown}{rgb}{0.4, 0.26, 0.13}
\definecolor{darksienna}{rgb}{0.24, 0.08, 0.08}
\definecolor{darkpowderblue}{rgb}{0.0, 0.2, 0.6}
\usepackage[colorlinks]{hyperref}
\hypersetup{linkcolor={darkpowderblue},citecolor={cyan},urlcolor={darksienna}}  
\usepackage{euscript,graphicx}

\usepackage{autobreak}
\usepackage{amsthm,dsfont,amsfonts,amsmath,amssymb,wasysym}
\usepackage{euscript,color,fontenc,textcomp,relsize}
\usepackage{bm,url,float}

\DeclareMathOperator{\sech}{sech}
\DeclareMathAlphabet\mathbfcal{OMS}{cmsy}{b}{n}

\allowdisplaybreaks

\allowdisplaybreaks

\begin{document}

\title{Gravitational Waves from Black-Hole Encounters: Prospects for Ground- and Galaxy-Based Observatories}
\author{Subhajit Dandapat}
\email{subhajit.phy97@gmail.com}
\affiliation{Department of Astronomy and Astrophysics,
 Tata Institute of Fundamental Research, Mumbai 400005, Maharashtra, India}
 
\author{Michael Ebersold}
 \email{michael.ebersold@lapp.in2p3.fr}
 \affiliation{Physik-Institut, Universit\"{a}t Z\"{u}rich, Winterthurerstrasse 190, 8057 Z\"{u}rich, Switzerland}
 \affiliation{Laboratoire d'Annecy de Physique des Particules, CNRS, 9 Chemin de Bellevue, 74941 Annecy, France}

\author{Abhimanyu Susobhanan}
\email{abhimanyu.susobhanan@nanograv.org}
 \affiliation{Department of Astronomy and Astrophysics,
 Tata Institute of Fundamental Research, Mumbai 400005, Maharashtra, India}
 \affiliation{National Centre for Radio Astrophysics, Tata Institute of Fundamental Research, Pune 411007, Maharashtra, India}
 \affiliation{Center for Gravitation Cosmology and Astrophysics, University of Wisconsin-Milwaukee, Milwaukee, WI 53211, USA}

 \author{Prerna Rana}
\affiliation{Department of Astronomy and Astrophysics,
 Tata Institute of Fundamental Research, Mumbai 400005, Maharashtra, India}

\author{A.~Gopakumar}
\affiliation{Department of Astronomy and Astrophysics,
 Tata Institute of Fundamental Research, Mumbai 400005, Maharashtra, India}
 
 \author{Shubhanshu Tiwari}
 \affiliation{Physik-Institut, Universit\"{a}t Z\"{u}rich, Winterthurerstrasse 190, 8057 Z\"{u}rich, Switzerland}

\author{Maria Haney}
 \affiliation{Nikhef, Science Park 105, 1098 XG Amsterdam, The Netherlands}
\author{Hyung Mok Lee}
\affiliation{
Center for the Gravitational-Wave Universe, Astronomy Program Department of Physics and Astronomy,
Seoul National University, 1 Gwanak-ro, Gwanak-gu, Seoul 08826, Korea
}

\author{Neel Kolhe}
 \affiliation{Department of Physics, St. Xavier’s College (Autonomous), Mumbai 400001, Maharashtra, India}

\begin{abstract}

Close hyperbolic encounters of black holes (BHs) generate certain Burst With Memory (BWM) events in the frequency windows of the operational, planned, and proposed  gravitational wave (GW) observatories.
We present  detailed explorations of the detectable parameter space of such events that are relevant 
for the LIGO-Virgo-KAGRA and the International Pulsar Timing Array (IPTA) consortia. 
The underlying temporally evolving GW polarization states are adapted from Cho et al.~[Phys.~Rev.~D 98,~024039~(2018)] and therefore incorporate general relativistic effects up to the third post-Newtonian order. 
Further, we provide a prescription to ensure the validity of our waveform family while 
describing close encounters.
Preliminary investigations reveal that optimally placed BWM events should be visible to megaparsec distances for the existing ground-based observatories.
In contrast, maturing IPTA datasets should be able to provide constraints on the occurrences of such hyperbolic encounters of supermassive BHs to gigaparsec distances.

\end{abstract}

\maketitle

\section{Introduction}
\label{sec:intro}
Observations of stellar mass compact binaries merging along quasi-circular orbits by the LIGO-Virgo-KAGRA collaboration,
numbering around 100,
have inaugurated the gravitational wave (GW) astronomy era \cite{abbott2021gwtc2,abbott2021gwtc3}.
This is mainly due to the rapid improvements in the sensitivities of the operational GW observatories \cite{LIGO,VIRGO}.
The maturing Pulsar Timing Array \cite[PTA:][]{foster1990constructing} experiments are expected to unveil the nanohertz (nHz) GW universe in the near future \cite{arzoumanian2020nanograv,goncharov2021evidence, chen2021common, Antoniadis2022iptagw}. 
In the coming decades, millihertz space-based GW observatories and third-generation ground-based and  decihertz GW observatories should allow us to pursue multi-band GW astronomy \cite{baker2019laser, amaro2022effect, gong2021concepts,maggiore2020science, jani2021gravitational, harms2021lunar}.

The existing, planned, and proposed ground-based GW observatories 
are expected to detect GWs from compact binaries in non-circular orbits \cite{abbott2019search, fragione2019eccentric, tagawa2021eccentric,chen2021observation,Ramos-Buades:2020eju,Ebersold:2022zvz}.
This includes relativistic hyperbolic encounters between black holes (BHs) and neutron stars (NSs) that manifest as GW Burst events \cite{mukherjee2021gravitational,kocsis2006detection, o2009gravitational, tsang2013shattering}.
Interestingly, millihertz GW observatories should be sensitive to such transient events that involve astrophysical and primordial BHs \cite{garcia2017gravitational}.
Further, PTAs could detect and characterize such GW burst events \cite{finn2010detection} after the eventual detection of a nHz GW background~\cite{burke2019astrophysics,taylor2021nanohertz}.

These considerations are prompting many detailed efforts that probe the feasibility of 
 such hyperbolic/parabolic encounters between BHs and NSs in astrophysically realistic simulations~\cite{rasskazov2019rate,tagawa2020formation,jaraba2021black}.
 In literature, 
the post-Newtonian (PN) approximation  is typically used to describe various aspects of GWs from hyperbolic encounters \cite{damour1985general,blanchet1989higher,junker1992binary}.
This approximation requires slow motion and weak fields, usually characterized by $({v}/{c})^2 \ll 1 $ and $ ( G\, M/ c^2 \, r ) \ll 1$ where $v$, $M$, and $r$ are respectively the orbital velocity, total mass, and relative separation of the binary \cite{blanchet2014gravitational}.  In contrast, 
 hyperbolic and parabolic encounters between BHs are also being described using the Effective One Body formalism, Numerical Relativity, and post-Minkowskian approaches \cite{nagar2021effective,gamba2021gw190521,nagar2021effectiveb,bae2020gravitational,khalil2022energetics}. 

This paper is structured as follows.
In Sec.~\ref{Sec2} , we 
provide details of a new GW template family that employs 
various inputs from Ref.~\cite{cho2018gravitational}  which will be useful for future searches of  GWs from compact binaries in hyperbolic orbits.
These events may be categorized as GW Burst signals that exhibit certain linear memory after the flybys \cite{devittori2014gravitational}.
We identify the regions of the parameter space in which we should expect hyperbolic encounters detectable by various types of GW observatories, influenced by \cite{bini2021frequency}.
Thereafter, we probe preliminary data implications of our approximant by estimating the distance reach of these events for the second and third-generation GW observatories.
In  Sec.~\ref{Sec3}, we provide the ready-to-use PTA response to GWs from PN-accurate hyperbolic passages of Supermassive black-hole binaries (SMBHs) and list the details of our \texttt{ENTERPRISE} \cite{ellis2019enterprise}- a compatible code that should be relevant for searching the resulting Burst With (linear) Memory events in the PTA datasets.

\section{PN-accurate approach to construct our BWM Waveform family}
\label{Sec2}

We begin by briefly describing our approach to construct temporally evolving quadrupolar order GW polarization states $h_{+,\times}$ associated with comparable mass compact binaries in fully 3PN-accurate hyperbolic orbits, and how we obtain the frequency content of our BWM waveform family. 
Thereafter, we provide a brief description of our  hyperbolic approximant and our estimates for the horizon distances of such events.

\subsection{Temporally Evolving Qudrupolar  \texorpdfstring{$h_{\times,+}(t)$} {hpx(t)} for Compact Binaries in PN-accurate Hyperbolic Orbits}

This subsection describes our PN-accurate approach to obtaining a GW template family for hyperbolic encounters.
We begin by displaying the  quadrupolar order GW polarization states $h_{+,\times|Q}$ associated with non-spinning compact binaries moving in non-circular orbits  characterized by the total mass $M=m_1+m_2$ and symmetric mass ratio $\eta=m_1m_2/M^2$ at 
 a luminosity distance $R'$.
\begin{widetext}
\begin{subequations}
\label{hcpQ}
\begin{align}
    h_{+|Q}&=-\frac{G M \eta}{c^4 \, R^\prime} \bigg{\{} (1+\cos^2i) \left[\left(\frac{G M}{r}+r^2\dot{\phi}^2-\dot{r}^2\right)\cos 2 \phi + 2 r \dot{r} \dot{\phi} \sin 2 \phi\right]
    +\sin^2 i \,  \left(\frac{G M}{r}-r^2 \dot{\phi}^2-\dot{r}^2\right)\bigg{\}} \\
    h_{\times|Q}&=-2 \, \frac{G M \eta}{c^4 \, R^\prime} \cos i \bigg{\{} \left(\frac{G M}{r}+r^2 \dot{\phi}^2-\dot{r}^2\right) \sin 2 \phi-2 r \dot{r}\dot{\phi} \cos 2 \phi \bigg{\}} \,,
\end{align}
\end{subequations}
\end{widetext}
where $i$ is the orbital inclination~\cite{cho2018gravitational}.
Further, the dynamical variables $r$, $\phi$, $\dot r$, and $\dot \phi$ stand for the radial and angular coordinates of the orbit in the center of mass frame and their time derivatives, respectively.
We model the temporal evolution of these dynamical variables during the hyperbolic encounters by employing a Keplerian-type parametric solution.
Specifically, we adapt the 3PN-accurate quasi-Keplerian parameterization for compact binaries in PN-accurate hyperbolic orbits derived in Ref.~\cite{cho2018gravitational}, which reads 
\begin{widetext}
\begin{subequations}
\label{orbitalp}
\begin{align}
r &= a_r \left ( e_r\,\cosh u  -1 \right )\,,\\
\frac{2\pi}{P}(t-t_0) & = e_t\sinh u -u+(\frac{f_{4t}}{c^4}+\frac{f_{6t}}{c^6})\, \nu+(\frac{g_{4t}}{c^4}+\frac{g_{6t}}{c^6})\sin \nu+\frac{h_{6t}}{c^6}\sin 2\,\nu+\frac{i_{6t}}{c^6}\sin 3\,\nu\,, \\
\frac{2\pi}{\Phi}(\phi-\phi_0) &= \nu+(\frac{f_{4\phi}}{c^4}+\frac{f_{6\phi}}{c^6})\sin 2\,\nu+(\frac{g_{4\phi}}{c^4}+\frac{g_{6\phi}}{c^6})
\sin 3\,\nu+\frac{h_{6\phi}}{c^6}\sin 4\,\nu+\frac{i_{6\phi}}{c^6}\sin 5\,\nu\,,
\end{align}
\end{subequations}
\end{widetext}
where 
$ \nu  = 2\,\arctan \biggl [ \sqrt{ \frac{e_{\phi} +1 }{e_{\phi} -1 }  } \, \tanh \frac{u}{2} \biggr ]$ and 
$u$ stands for the eccentric anomaly, while  $a_r$, $e_r$, $e_t$, $n$, and $t_0$ are certain PN-accurate semi-major axis, radial eccentricity, time eccentricity, mean motion, and initial epoch, respectively.
Explicit 3PN-accurate expressions for orbital elements such as $P=2\pi/n$, $a_r$, $e_r$, $e_t$, $e_\phi$, and $\Phi$, as well as functions appearing in the generalized quasi-Keplerian parametrization such as $f_{4t}$, $g_{4t}$, $f_{4\phi}$, $g_{4 \phi}$, etc., in terms of the conserved energy $E$ and angular momentum $L$,  were derived in Ref.~\cite{cho2018gravitational}.

It is fairly straightforward to obtain a 3PN-accurate expression for $r$ and $\phi$  in terms of $E$, $L$, and $u$ while using the following expressions to obtain 3PN-accurate expressions for $\dot r$ and $\dot \phi$:
\begin{subequations}
	\begin{align}
	\frac{dt}{du}&=\frac{\partial t}{\partial u}+\frac{\partial t}{\partial \nu} \, \frac{d \nu}{d u} \,,\\[1ex]
 
	\dot{r}&=\left(\frac{dr}{du} \bigg{/} \frac{dt}{du}\right)\,, \\[1ex]
	\dot{\phi}&=\left(\frac{d\phi}{d\nu} \, \frac{d\nu}{du} \bigg{/}\frac{dt}{du}\right)\,.
	\end{align}
\end{subequations}

It is convenient to express the 3PN-accurate expressions of $r$, $\phi$,  $\dot{r}$, and $\dot{\phi}$ in terms of the dimensionless parameter $x\equiv \left( \frac{G\,M\,n}{c^3}\right)^{2/3}$ (where $n={2 \pi}/P$ is defined in Eq. (2.36c) of \cite{cho2018gravitational}), the time eccentricity $e_t$, and the eccentric anomaly $u$.
We employ the following 3PN-accurate expressions for $E$ and $h= \frac{L}{ G\,M^2\, \eta} $ in terms of $x$ and $e_t$ in modified harmonic coordinates that can be extracted from Ref.~\cite{cho2018gravitational}:
%
\begin{widetext}
\begin{subequations}
\label{Eh_3PN}
\begin{align}
\frac{1}{c^2 h^2}&=\frac{x}{e_t^2-1}+\frac{x^2}{3 \, (e_t^2-1)^2} \bigg{\{}-3+e_t^2 (9-5 \eta )-\eta \bigg{\}}+\frac{x^3}{12 \, (e_t^2-1)^3}  \bigg{\{}-60+27 \eta \notag \\
&+e_t^4 \bigg{(}48-17 \eta +20 \eta ^2\bigg{)}+e_t^2 \bigg{(}-36+62 \eta +28 \eta ^2\bigg{)}\bigg{\}}-\frac{x^4}{(e_t^2-1)^4} \bigg{\{} \frac{32}{3}+\frac{1}{96} \bigg{(}-4124+123 \pi ^2\bigg{)} \eta \notag \\
&+\frac{71 \eta ^2}{36}-\frac{\eta ^3}{81}-e_t^4 \bigg{(}-6+\frac{563
	\eta }{8}-\frac{1249 \eta ^2}{36}-\frac{149 \eta ^3}{27}\bigg{)}-e_t^2 \bigg{(}-89+\bigg{(}\frac{57193}{280}-\frac{123 \pi ^2}{32}\bigg{)} \eta  \notag \\
&-\frac{1465
	\eta ^2}{36}-\frac{34 \eta ^3}{27}\bigg{)}+\frac{1}{648} e_t^6 \bigg{(}-1080+27 \eta +414 \eta ^2+800 \eta ^3 \bigg{)}\bigg{\}} \,,\\
\frac{2E}{c^2}&=x+\frac{x^2}{12} (\eta -15)+\frac{x^3}{24} (15 -15 \, \eta-\eta^2 )+\frac{5 \, x^4}{5184} \bigg{(} 999+1215 \eta +90 \eta^2+7 \eta^3\bigg{)}.
\end{align}
\end{subequations}
\end{widetext}
\par

In what follows, we list the fully 1PN-accurate expressions of $r$, $\dot{r}$, $\phi$, and $\dot{\phi}$ in terms of $x$, $e_t$, and $u$ to demonstrate the structure of these expressions. 
\begin{widetext}
\begin{subequations}
\label{orbit_1}
\begin{align}
r(u)&=\frac{GM}{c^2} \left\{ \frac{e_t \cosh u-1}{x} +\frac{1}{6} \, \bigg{(} 2(\eta-9)+e_t \, (7 \eta -6) \, \cosh u \bigg{)}\right\}\,, \\
\dot{r}(u)&=\frac{c \, e_t \, \sqrt{x} \sinh u}{e_t \cosh u-1}\left\{1+x \, \frac{7 \eta-6}{6} \right\}\,, \\
\phi(u)-\phi_0&=2 \arctan\left[ \sqrt{\frac{e_\phi+1}{e_\phi-1}} \, \tanh \frac{u}{2}\right] \, \left\{ 1+\frac{3 \, x}{e_t^2-1}\right\} \,,\\
\dot{\phi}(u)&=\frac{c^3}{G \, M} \Bigg{\{} \frac{\sqrt{e_t^2-1} \, x^{3/2}}{(e_t \cosh u -1)^2}-\frac{x^{5/2}(3+e_t^2(\eta-4)+e_t(1-\eta)\cosh u)}{\sqrt{e_t^2-1}(e_t \cosh u-1)^3}\Bigg{\}}\,.
\end{align}
\end{subequations}
\end{widetext}
We emphasize that we have employed the 3PN version of these expressions in our GW template family.
In practice, we employ the following 3PN-accurate expression to express $ \sqrt{\frac{e_\phi+1}{e_\phi-1}} $ in terms of $e_t$ and $x$ while describing the angular motion:
\begin{widetext}
\begin{align}
\label{eq:ephiet}
 \sqrt{\frac{e_\phi+1}{e_\phi-1}}&=\sqrt{\frac{e_t+1}{e_t-1}} \Bigg{[}1+\frac{x}{e_t^2-1} \bigg{\{} e_t (4-\eta ) \bigg{\}} +\frac{x^2}{(e_t^2-1)^2} \bigg{\{} e_t \bigg{(}21-\frac{65 \eta }{24}-\frac{\eta ^2}{24}\bigg{)}+e_t^2 \bigg{(}8-4 \eta +\frac{\eta ^2}{2}\bigg{)} \notag \\
&+e_t^3 \bigg{(}4-\frac{109
	\eta }{96}+\frac{55 \eta ^2}{96}\bigg{)}\bigg{\}}+\frac{x^3}{(e_t^2-1)^3}\bigg{\{} e_t \bigg{(}154-\frac{44687 \eta }{336}+\frac{41 \pi ^2 \eta }{64}-\frac{139 \eta ^2}{24}\bigg{)}+e_t^4 \bigg{(}16 \notag \\
&-\frac{205 \eta
}{24}+\frac{329 \eta ^2}{96}-\frac{55 \eta ^3}{96}\bigg{)}+e_t^5 \bigg{(}\frac{213 \eta }{128}-\frac{61 \eta ^2}{384}-\frac{71 \eta ^3}{384}\bigg{)}+e_t^2
\bigg{(}84-\frac{191 \eta }{6}+\frac{61 \eta ^2}{24}+\frac{\eta ^3}{24}\bigg{)}\notag \\
&+e_t^3 \bigg{(}110-\frac{16663 \eta }{420}-\frac{41 \pi ^2 \eta }{256}-\frac{1325 \eta ^2}{192}+\frac{55 \eta ^3}{192}\bigg{)}\bigg{\}}\Bigg{]}\,.
\end{align}
\end{widetext}

We are now positioned to incorporate the effects of GW emission that enters the orbital dynamics at 2.5PN (absolute) order. 
This is achieved by adapting the GW phasing formalism, developed for eccentric inspirals in Ref.~\cite{damour2004phasing}.
The plan involves computing first the time derivatives of Newtonian expressions for $n= 2\,\pi/P$ and $e_t^2$ in terms of the conserved orbital energy and angular momentum. 
Thereafter, we replace the time derivatives of $E$ and $L$ with the 2.5PN-accurate (absolute) far-zone energy and angular momentum flux expressions, given  in Ref.~\cite{blanchet2014gravitational}.
We now replace the variables $r$, $\dot r$, $\dot \phi$, $E$, and $L$, present in the $dn/dt$ and $de_t/dt$ expressions, by their Newtonian counterparts obtained from Eqs.~(\ref{Eh_3PN}) and (\ref{orbit_1}).
This leads to the following quadrupolar order expressions for $dx/dt$ and $d e_t/dt$ in the modified harmonic gauge: 
\begin{subequations}
\label{gw_em}
\begin{align}
\frac{d x}{dt} &=\frac{16}{15} \, \frac{c^3 \, x^5 \, \eta}{G \, M \, \beta^6} 	 \bigg{\{} 35 \, (1-e_t^2)+(49 -9 e_t^2)\, \beta+32 \, \beta^2 \notag \\
&+ 6\, \beta^3 \bigg{\}}
\,,\\
\frac{d e_t}{dt} &= \frac{8}{15} \, \frac{(e_t^2-1) \, x^4 \, c^3\, \eta}{G \, M \, e_t \, \beta^6} \bigg{\{} 35 \,  (1-e_t^2)+(49-9 \, e_t^2) \, \beta \notag \\
&+17 \, \beta^2+3 \beta^3 \bigg{\}}
\,,    
\end{align}
\end{subequations}
where $\beta=(e_t \, \cosh u-1)$. 
It should be obvious that the evolution equations for $x$ and $e_t$ depend on the variables $x$, $e_t$, and $u$.
 
A close inspection of these expressions reveals that we are now in a position to obtain  $h_{+,\times|Q}$ as a function of  $u$ for hyperbolic encounters of compact binaries, characterized by  $M$, $\eta$, $n$, and $e_t$ by employing 3PN-accurate expressions for $r$, $\dot r$, $\phi$, and $\dot \phi$ in Eqs.~\eqref{orbit_1} while solving the above-given coupled differential equations for $x$ and $e_t$ for incorporating the effects of GW emission.
However, we would like to have $h_{+,\times|Q}$ as a function of time and we provide the following 3PN-accurate differential equation for $u$ that can be  extracted from the 3PN-accurate Kepler equation, given by Eq.~(\textcolor{darkpowderblue}{2.35}) in Ref.~\cite{cho2018gravitational}, 
the 2PN version of the relevant equation reads (the 3PN-accurate version is provided in Appendix \ref{dudt3pn}).
 \begin{align}
 \label{dudt}
\frac{du}{dt} &=\frac{x^{3/2} \, c^3}{G \, M \, \beta} \bigg{\{} 1-\frac{x^2}{8 \, \beta^3} \bigg{[} (60-24 \, \eta)\, \beta+(15-\eta) \notag \\
& \quad\times\eta \, e_t \, (e_t-\cosh u) \bigg{]} \bigg{\}}
\,,
 \end{align}
where $\beta=(e_t \, \cosh u-1)$ as before. 
 
To obtain temporally evolving $h_{+,\times|Q}$ associated with compact binaries in fully 3PN-accurate hyperbolic orbits, we pursue the following steps.
First, we specify the initial values for $e_t$, $n$, and $u$ for a compact binary that is characterized by $M$ and $\eta$.
Thereafter, we solve numerically the above-listed differential equations for $n$, $e_t$, and $u$ to track the temporal evolution of these variables.
This naturally leads to the PN-accurate temporal evolution of our dynamical variables, namely $r$, $\dot r$, $\phi$, and $\dot \phi$. 
It is now straightforward to obtain $h_{+,\times|Q}(t)$ associated with compact binaries in  fully 3PN-accurate hyperbolic orbits with the help of Eqs.~(\ref{hcpQ}).
  
In the next subsection, we explain how to adapt the present prescription to model BWM events in the distinct frequency windows of various types of GW observatories.

\subsection{Characterizing BWM events }

This section tackles two points that will be relevant while probing data analysis implications of our time-domain GW signal for various GW observatories.
First, it should be obvious that the present prescription does not reveal the frequency content of these GW events. 
Secondly, our $h_{+,\times|Q}(t)$ family requires us to specify $n$ which is not a commonly used parameter to characterize hyperbolic encounters.
We first tackle this issue by introducing a PN-accurate impact parameter $b$ and expressing it in terms of $n$ and $e_t$. 
Influenced by Ref.~\cite{blanchet1989higher}, we define a ~PN-accurate impact parameter $b$ such that $b\, \text{v}_{\infty} = | \bm r \times \textbf{ v} |$ when $ |\bm r | \rightarrow \infty$, where $\text{v}_{\infty}$ stands for the relative velocity at infinity.
 It is now straightforward to obtain the 3PN-accurate expression for $b$  in terms of $x$ and $e_t$  as presented in Ref~\cite{cho2018gravitational}. 
We display the relevant expression for $b$ in terms of $x$ and $e_t$ in the modified harmonic gauge as
\begin{align}\label{b_hyp}
b=& \zeta \, \frac{\sqrt{e_{\rm t}^2-1}}{ x}\Bigg\{\,1- x \, \left(\frac{\eta-1}{e_{\rm t}^2-1}+\frac{7\eta-6}{6}\right) \notag \\[2ex]
&+x^2 \, \bigg[1-\frac{7}{24}\eta+\frac{35}{72}\eta^2+\frac{3-16\eta}{2(e_t^2-1)}+\frac{7-12\eta-\eta^2}{2(e_t^2-1)^2}\bigg]\notag \\[2ex]
&+x^3\,\bigg[-\frac{2}{3}+\frac{87}{16}\eta-\frac{437}{144}\eta^2+\frac{49}{1296}\eta^3+\notag \\[2ex]
&+\frac{36-378\eta+140\eta^2+3\eta^3}{24(e_t^2-1)}+\frac{1}{6720(e_t^2-1)^2}\big\{248640 \notag \\[2ex]
&+(-880496+12915\pi^2)\,\eta+40880\,\eta^2+3920\,\eta^3\big\}\notag \\[2ex]
&+\frac{1}{1680(e_t^2-1)^3}\big\{73080+(-228944+4305\pi^2)\eta\\\notag
&+47880\eta^2+840\eta^3\big\}\bigg]\Bigg\}~,
\end{align}
where $\zeta=GM/c^2$. 
It should be obvious that we can now characterize compact binaries in PN-accurate hyperbolic orbits with the help of $m_1$, $m_2$, $e_t$, and $b$.


We now proceed to address the frequency content of our time-domain GW burst events.
This is done by looking into the quadrupolar order expression for the total energy radiated during hyperbolic encounters. 
It is convenient to pursue such a calculation in the time domain by employing quadrupolar order GW energy flux expression for compact binaries in non-circular orbits and the Keplerian type parametric solution, as detailed in Ref.~\cite{blanchet1989higher}.
In other words, the Newtonian estimate for radiated energy during the hyperbolic encounter may be written as 
\begin{align}
\Delta \mathcal{E_\text{Q}}=\int^{+\infty}_{-\infty} dt \, \mathcal{F'_\text{Q}}(t)=\int^{+\infty}_{-\infty} du \, \left(\frac{dt}{du}\right) \mathcal{F'}_\text{Q}\,,
\end{align}
where $ \mathcal{F'_\text{Q}}(t)$ stands for the quadrupolar order GW luminosity expressed in terms of $r$, $\dot r$, and $\dot \phi$, given by Eq.~(3.41) in Ref.~\cite{blanchet1989higher}.
Additionally, $\mathcal{F'_\text{Q}}$ can be expressed in terms of $e_t$, $n$, and $u$ as given by Eq.~(5.7) in Ref.~\cite{blanchet1989higher}.
This leads to 
\begin{align}
\label{E_q}
\Delta \mathcal{E_\text{Q}}&=
\frac{2 M \, \eta ^2}{15 c^5 \, h^7} 
\, \bigg{[}\sqrt{e^2-1} \bigg{(}\frac{602}{3}+\frac{673 \,  e^2}{3}\bigg{)} \nonumber \\
&+\bigg{(}96+292 \, e^2+37 \, e^4\bigg{)} \arccos \bigg{(}-\frac{1}{e}\bigg{)} \bigg{]}\,,
\end{align}
where $e$ is the Newtonian eccentricity \cite{blanchet1989higher}.
Very recently, Ref.~\cite{cho2021instantaneous} provided a 3PN version of the above result that extended the 1PN-accurate result of Ref.~\cite{blanchet1989higher}.

However, it is possible to obtain a similar estimate while pursuing the computation in the frequency domain, as detailed in Ref.~\cite{bini2021frequency}.
The relevant expression reads 
\begin{align}
\Delta \mathcal{E_\text{Q}}  &= \int_0^\infty d \omega \, F_{\text Q}(\omega)\,,
\end{align}
where $F_{\text Q}(\omega)$ stands for the Fourier domain version of the GW luminosity. 
For the present investigation, we employ the Newtonian accurate Fourier domain expression for the GW luminosity, given by Eqs. (3.27) in \cite{bini2021frequency} and it reads
\begin{widetext}
\begin{align}
\label{fourierFw}
 F_{\text Q}(\omega)   
&=\frac{32}{5} \, \frac{G}{\pi \, c^5} \eta^2 \left( \frac{G M^2}{a\, z \, e}\right)^2 \, \text{e}^{\pi z/e} \bigg{\{} z^2 (p^2+z^2+1)(p^2+z^2)K_{p+1}^2(z)-2 \ z \bigg{[} \left( p-\frac{3}{2}\right)  z^2+p(p-1)^2 \bigg{]} \notag \, \\[2ex]
&\times (p^2+z^2)K_p(z) K_{p+1}(z)+2 \bigg{[} \frac{z^6}{2}+\bigg{(}2p^2-\frac{3}{2}p+\frac{1}{6}\bigg{)}z^4+\bigg{(} \frac{5}{2}p^4-\frac{7}{2}p^3+p^2 \bigg{)}z^2+p^4(p-1)^2\bigg{]}K_p^2(z)\bigg{\}}, 
\end{align}
\end{widetext}
where $z$ and $p$ are dimensionless parameters given by $z=\frac{\omega \, e \, a^{3/2}}{\sqrt{G\, M}}$, and $p=\frac{i \, z}{e}$. Note that, here $i$ refers to the imaginary number and it should not be confused with the orbital inclination defined in Eq.~\eqref{hcpQ}.
In the above expression, $e$ and $a$ denote the Newtonian eccentricity and semi-major axis, respectively. 
Further,  we require the relation that connects the semi-major axis with the impact parameter $b$ at Newtonian order, namely $a=\frac{b}{\sqrt{e^2-1}}$.
It should be noted that the quadrupolar expression requires both the total mass and the mass ratio and can not be written only in terms of the chirp mass $M_c =M \, \eta^{3/5}$. 
However, the peak frequency of emitted GW will be independent of $\eta$, since $\eta$ only appears as an overall multiplicative factor in the Fourier domain luminosity expression as evident from Eq.~\eqref{fourierFw}.

\begin{figure*}
\includegraphics[width=0.85\linewidth]{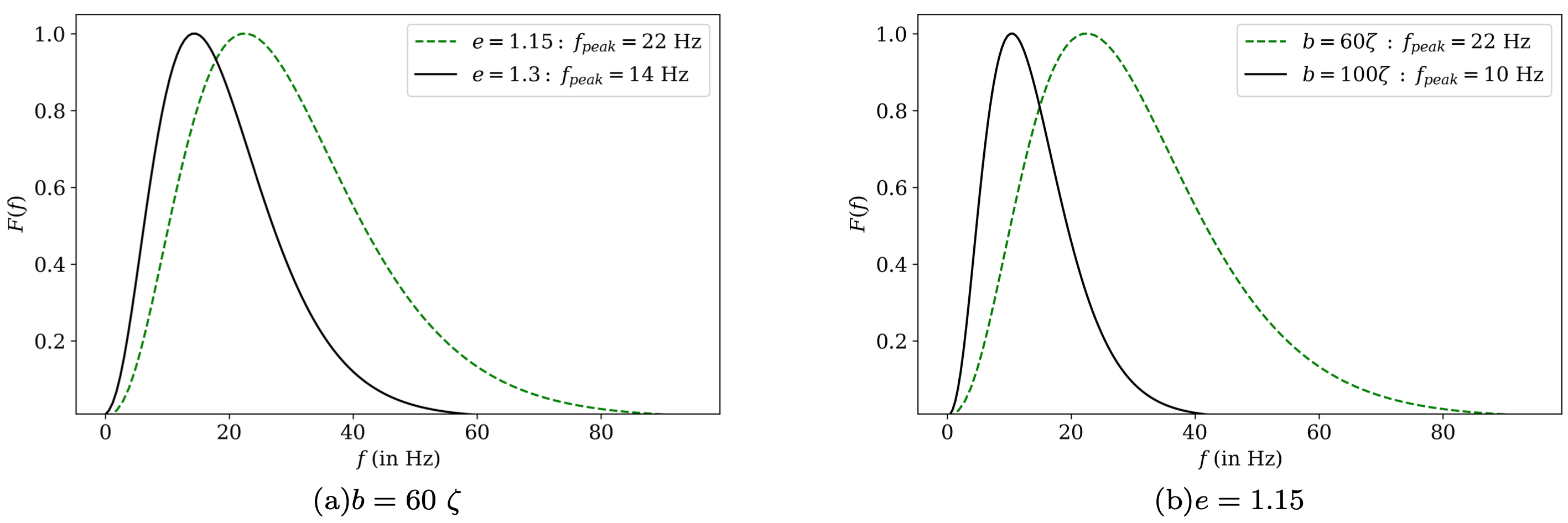}
\caption{ Quadrupolar order GW power spectrum  from hyperbolic encounters.
We let $M=40 M_\odot$ while varying  the orbital eccentricity (left figure) and the impact parameter (right figure). We infer that higher $e_t$ and $b$ values lead to narrow-band signals.
Note that we are plotting  the normalized $F(f)$  using the power associated with the peak frequencies of the associated distribution. We use $f=\omega/(2 \pi)$ while listing the peak frequencies and $\zeta = G\,M/c^2$.
}
\label{figbe}
\end{figure*}

\begin{figure*}
\includegraphics[width=0.85\linewidth]{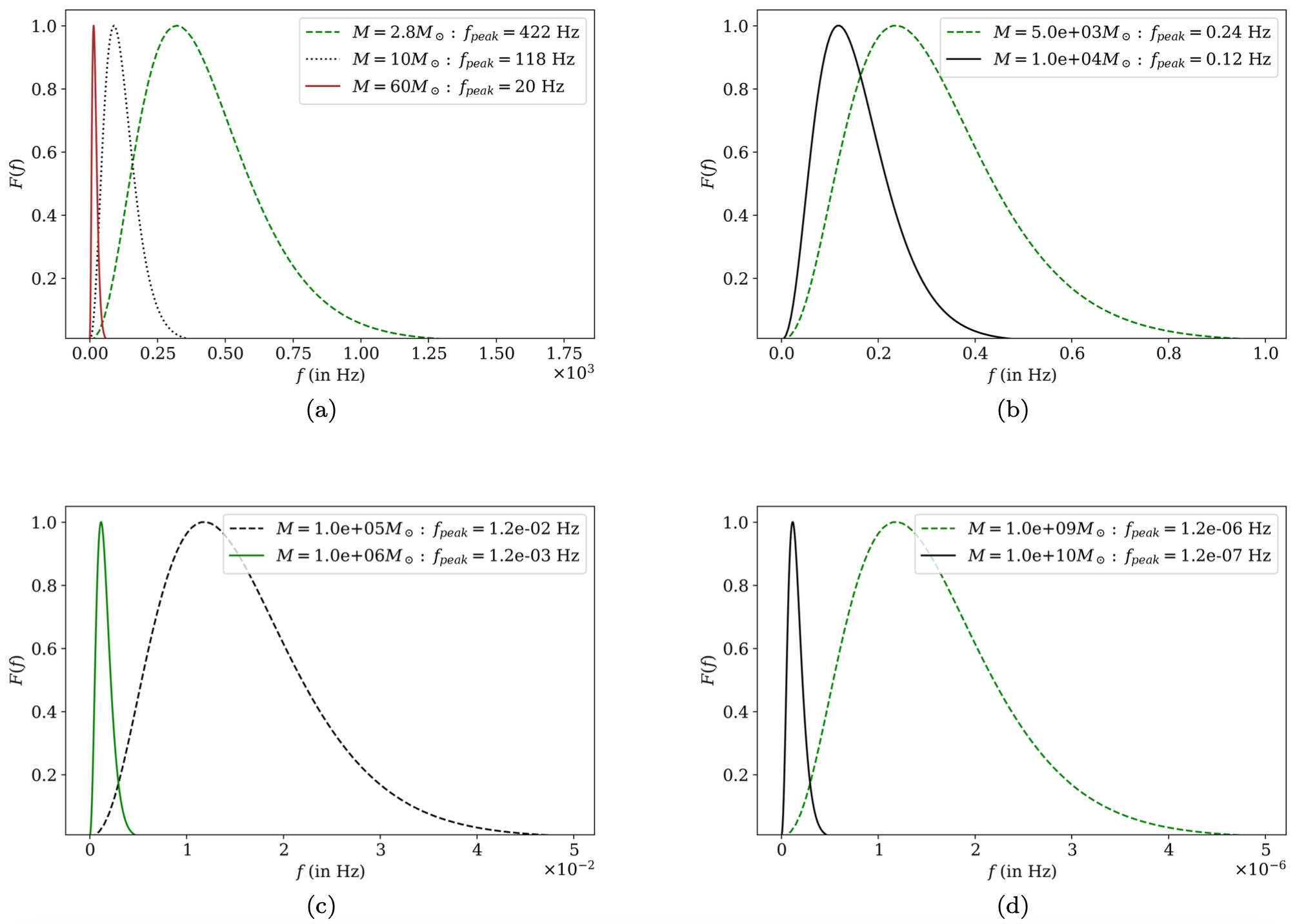}
\caption{ Quadrupolar order GW spectra from hyperbolic encounters that are specified by an impact parameter ($b=60 \zeta$) and eccentricity ($e=1.15$) while we vary  the total mass $M$.
These plots reveal that hyperbolic events can emit GWs in the LIGO, DECIGO~\cite{kawamura2019space}, LISA, and PTA frequency windows.
From the listed  $f_\text{peak} $  values, we infer its $ \sim 1/M$  dependency and these plots also reveal that lower $M$ systems provide broad-band spectra compared to higher total mass systems. 
Note that the listed  GW frequencies $f=\omega/(2 \pi )$ are in Hertz and we normalize $F(f)$ using the power associated with the peak frequencies of the associated distribution.
}
\label{fig_m}
\end{figure*}

We now employ the above quadrupolar order expression to estimate the frequency spectrum of our $h_{+,\times|Q}(t)$ associated with compact binaries in PN-accurate hyperbolic orbits.
This is influenced by the way the GW frequency spectrum of eccentric binaries was detailed in Sec.~III of  Ref.~\cite{peters1963gwspectrum}. 
In order to obtain the peak frequency of the emitted GWs, we need to maximize Eq.~\eqref{fourierFw} with respect to $z$ and get the corresponding $\omega$ with the help of the aforementioned relation which connects $z$ to $\omega$.
Further, the peak frequency is inferred via $f_\text{peak}=\frac{\omega_\text{peak}}{2 \, \pi}$ and we note that $f_\text{peak}\sim \frac{c^3}{G\, M}$ for a fixed eccentricity and impact parameter value.
With the help of these inputs, we now show that our BWM waveform family can provide transient GW events in the frequency windows of Earth-, Solar System-, and Galaxy-based GW observatories. 
In Fig.~\ref{figbe}, we plot the GW energy spectrum with fixed total mass ($M=40 M_\odot$) while varying eccentricity and impact parameters.
In contrast, we plot $F_{\text Q}(\omega)$ for hyperbolic encounters, specified by $b= 60 \, \zeta$ and $e=1.15$ in Fig.~\ref{fig_m}, while varying the total mass of our fiducial equal mass BH binary. 

From these figures, we conclude that the lower the total mass, impact parameter, and eccentricity of the binary system, the broader the spectrum. 
Further, we infer that the stellar mass BH binaries can provide such transient events for the operational and planned ground-based GW observatories. 
However, DECIGO, LISA, and PTA relevant sources involve BH binaries that weigh thousands, millions, and billions of solar masses respectively and this is consistent with our observation that $f_\text{peak}\sim \frac{c^3}{G\, M}$.
The following caveat is worth mentioning:
strictly speaking, we should have employed the PN-accurate expression for $F(\omega)$, influenced by Ref.~\cite{tessmer2007accurate} that explored the effect of periastron advance on the GW spectrum of compact binaries in PN-accurate eccentric orbits.
Unfortunately, it is rather difficult to obtain closed-form expressions for the PN-accurate version of $F_{\text{Q}} (\omega)$ as detailed in Ref.~\cite{bini2021frequency}. 
We plan to tackle the PN-accurate extension of the present investigation in future work. 
However, we do not expect that PN corrections will substantially change the shape of $F_{\text{Q}} (\omega)$ and  the present $f_\text{peak}$  estimates.



The plots in Figs.~\ref{figbe} and \ref{fig_m} suggest that hyperbolic events, characterized by certain $(b,e_t, M, \eta)$ values, are potential GW sources for various types of GW observatories.
Therefore, it is important to ensure that the underlying PN approximation should be appropriate to describe these events.
Ideally, this should involve detailed comparisons with Numerical Relativity (NR) efforts involving BHs in hyperbolic orbits.
In the absence of such efforts, we restrict our attention to those  $(b,e_t, M, \eta)$ values that ensure that the orbital separation at the closest approach, namely 
$r_\text{min} \gtrsim 10 \, \zeta$.
This restriction, while somewhat arbitrary, is influenced by the fact that NR and PN descriptions agree rather nicely  with each other at such orbital separations while dealing with eccentric and circular  binaries
\citep{hinder2018eccentric, will2011unreasonable}.
Further, we have demonstrated that certain versions of PN-accurate hyperbolic fluxes are excellent approximations of GW fluxes from 
BH binaries 
that support high-bound eccentricities
at such orbital separations
\cite{cho2021instantaneous}.

\begin{figure}
\centering
\includegraphics[width=1\linewidth]{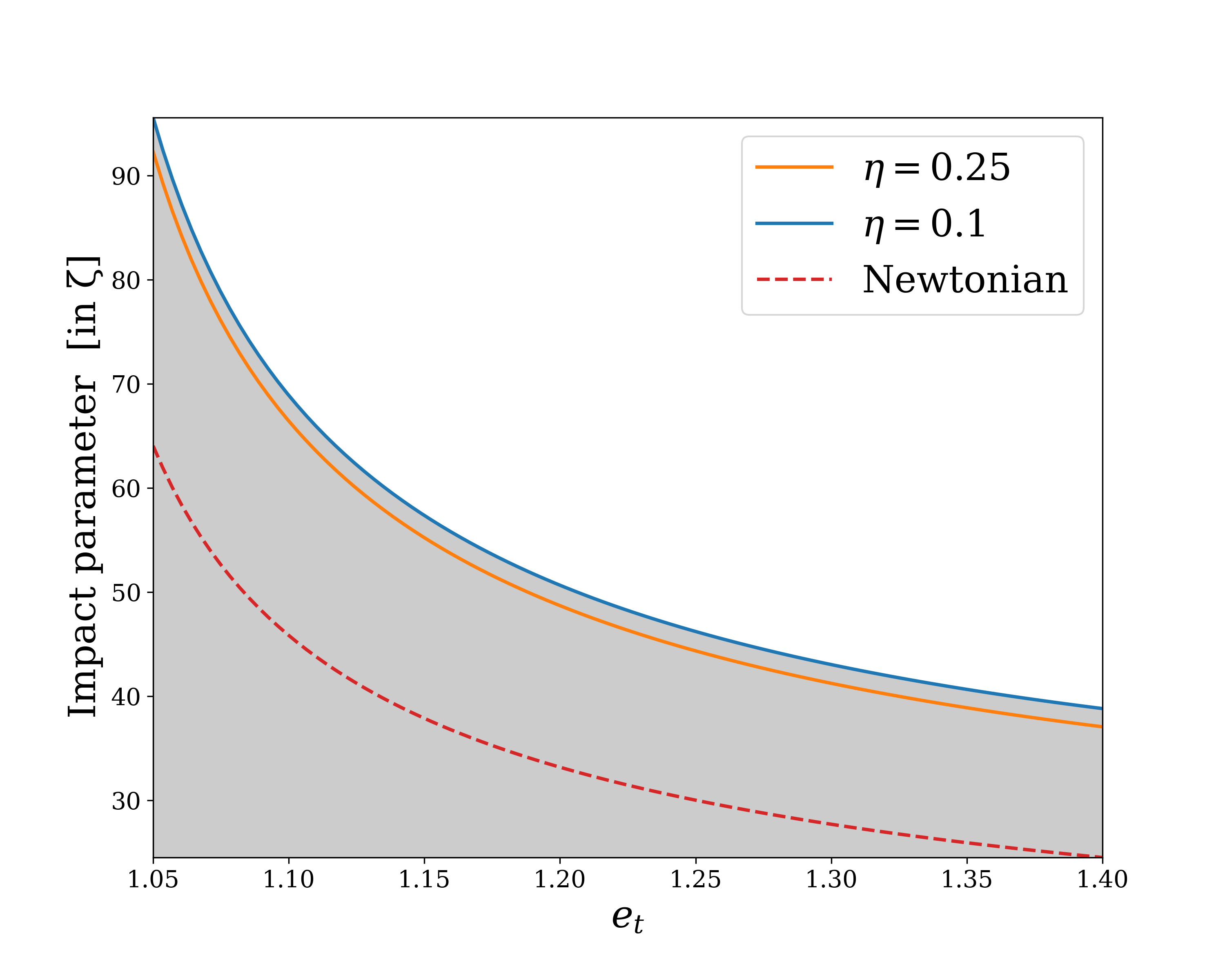}
\caption{
The  results of imposing $r_\text{min}= 10 \, \zeta $ on the     $(b,e_t)$ parameter space. 
For the present effort, we 
will not consider hyperbolic events with $b$ and $e_t$ values that are in the shaded region. 
Interestingly, the PN approximation provides     tighter restrictions compared to its Newtonian counterpart, and $\eta$ influences are rather minimal.
We find that both analytical and numerical approaches to impose     $r_\text{min}= 10 \, \zeta$ restriction provide    similar plots.
}
\label{fig:b_et_AR}
\end{figure}

We adopt the following approach to impose the restriction that  $r_\text{min} \gtrsim 10 \, \zeta$ while choosing $(b,e_t)$ values.
We begin with  the 3PN accurate expression of $r$ in terms of $x$, $e_t$, $\eta$, and $u$ and employ the 3PN accurate equation that connects $x$ to $(b, e_t, u)$ to obtain the 3PN accurate $r$ expression  in terms of $b$, $e_t$, $\eta$, and $u$. 
Thereafter, we invert the resulting expression with $r=r_\text{min}$ and $u=0$ to get the 3PN accurate expression of $b$ in terms of $\eta, r_\text{min}$ and $\eta$. 
The choice of  $u=0$ is natural as it provides the periastron point ($\phi=0$) in the center-of-mass  frame. 
The resulting 3PN accurate expression of $b$ in terms of $e_t, r_\text{min}$ and $\eta$ reads
\begin{widetext}
\begin{align}
	b&=\sqrt{\frac{e_t+1}{e_t-1}} \, r_\text{min}- \zeta \bigg{\{}-18+\eta +3 e_t (-8+3 \eta )+2 e_t^2 (-6+7 \eta )\bigg{\}}+\frac{1}{24
		(1+e_t) \sqrt{e_t^2-1}} \frac{\zeta^2}{r_\text{min}} \notag \\[2ex]
	& \bigg{\{}318-155 \eta +3 e_t^3 (29-3 \eta ) \eta +3 \eta ^2+3 e_t (32-85 \eta +3 \eta ^2)-e_t^2
	(216+297 \eta +5 \eta ^2) \notag \\[2ex]
	&+2 e_t^4 (195-218 \eta +55 \eta ^2)\bigg{\}}-\frac{\zeta^3}{20160 \, (1+e_t)^3 \, r_\text{min}^2} \,\sqrt{\frac{e_t+1}{e_t-1}} \,   \bigg{\{}-221760+(2330844 \notag \\[2ex]
	&-77490 \pi ^2) \eta -660940 \eta ^2 +1260 \eta ^3-420 e_t^5 (-3120+4229 \eta -1828 \eta ^2+306 \eta ^3)+14 e_t^2 (\notag \\[2ex]
	&-187200+(323412-9225 \, 
	\pi ^2) \eta +1000 \eta ^2 +380 \eta ^3)+280 e_t^6 (-4164+5106 \eta -2700 \eta ^2+565 \eta ^3)\notag \\[2ex]
	&-3 e_t^3 (322560+(-58328+4305
	\pi ^2) \eta -102760 \eta ^2+840 \eta ^3)+140 e_t^4 (15840-5367 \eta -2003 \eta ^2\notag \\[2ex]
	&+1139 \eta ^3)-3 e_t (-275520+(44012 +12915
	\pi ^2) \eta -51240 \eta ^2+1680 \eta ^3)\bigg{\}},
\end{align}
\end{widetext}
with $\zeta=\frac{G M}{c^2}$.
The resulting allowed regions of the $(b,e_t)$ parameter space are displayed in Fig.~\ref{fig:b_et_AR}.
It seems reasonable to choose $e_t$ values to be around $1.15$ when $b$ estimates are around $60\zeta$.
Further, lower $b$ and $e_t$ values can give interesting GW events in the Advanced LIGO (aLIGO) frequency window and it will be interesting to take a look at such events after pursuing proper PN versus NR comparisons that deal with BHs in hyperbolic orbits.
Additionally, we have  verified that a similar figure is obtained while numerically imposing our $r_\text{min}$ restriction in the 3PN-accurate expressions for $b$ and $r$, given by ~Eq.~\eqref{b_hyp} and \eqref{orbit_1}.
This provides additional assurance for the validity of PN approximation in these hyperbolic orbits.


\begin{figure*}
\includegraphics[width=0.85\linewidth]{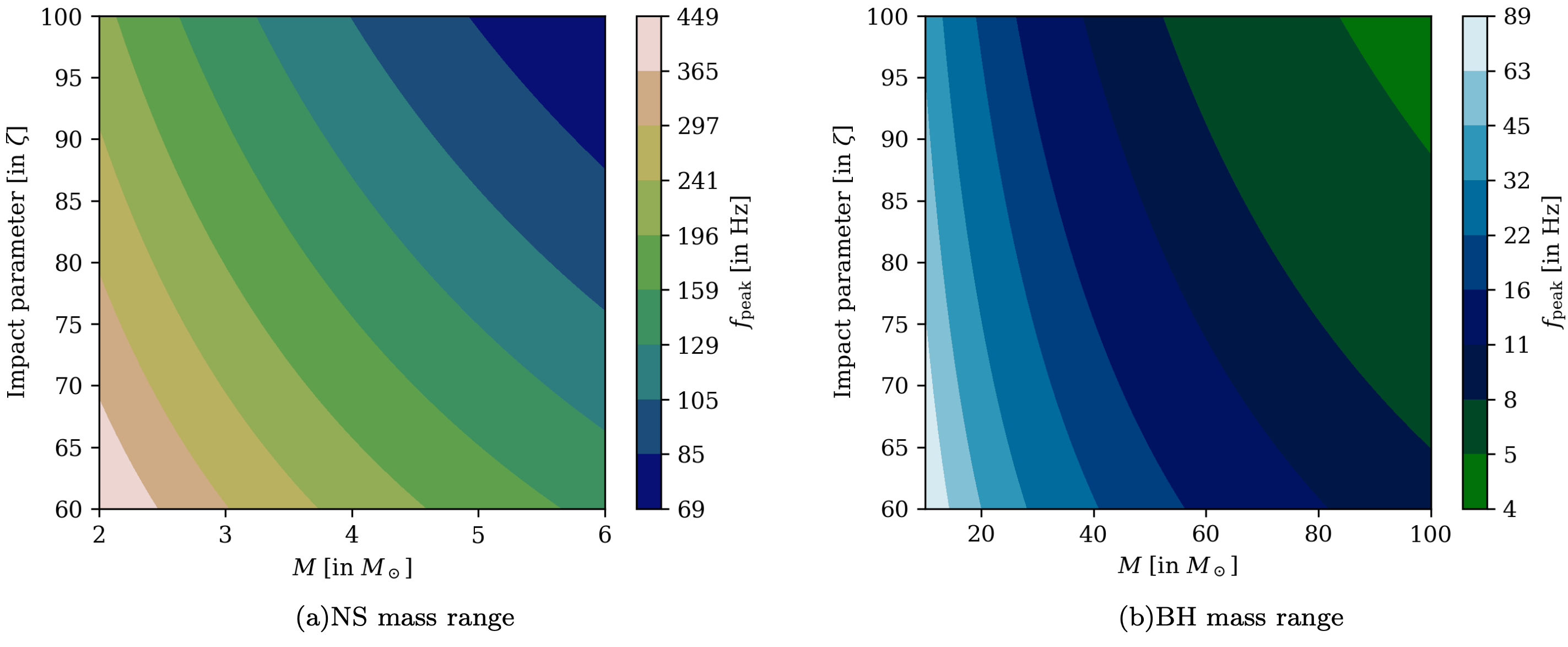}
\caption{ Contour plots that the probe peak frequencies of LVK-relevant hyperbolic encounters for equal mass Binary Neutron Star (BNS) and Binary Black Hole (BBH) systems with fixed eccentricity e=1.15. We have considered the system with a total mass of $M \gtrsim 6 M_\odot$ as BBH~\cite{rhoades1974maximum}. Lower $b$ and $e_t$ values can lead to higher GW frequency events while lower $\eta$ systems provide lower GW amplitudes. 
These plots only indicate approximate regions of the LVK relevant parameter space for such encounters due to the broadband nature of the resulting GWs.}
\label{peakfreq-heatmap}
\end{figure*}

We now try to specify the region of  $(M,b,e_t)$ parameter space that should be relevant to GW observatories like the aLIGO and the planned Einstein Telescope (ET) \cite{maggiore2020science}. 
For this purpose, we compute the peak frequencies for  equal mass hyperbolic encounters  as functions of the  total mass and the impact parameter while  letting $e_t = 1.15$.
We have taken additional precautions to ensure that the resulting events  can be accurately described by the PN approximation as discussed earlier.
Our results are displayed in Fig.~\ref{peakfreq-heatmap} while considering hyperbolic events involving  neutron stars  with masses up to 2-6~$M_\odot$ and stellar-mass black holes in the mass range of 10-100~$M_{\odot}$.
These plots reveal that higher  $M$ and $b$ values lead to lower peak frequencies. 
However,  higher impact parameters also lead to a decrease in GW amplitudes and these considerations indicate that stellar-mass BHs and encounters of neutron stars are the most interesting source for ground-based detectors. 
It turned out that lower $b$ and $e_t$ values can provide higher GW frequency events though further investigations will be required to substantiate the use of PN approximation to describe such events.
We would like to note that the tidal interactions should not play any significant role in our hyperbolic events involving neutron stars.
This is mainly because such interactions are expected to occur at the 5PN order which is beyond the accuracy of our description \cite{Bernuzzi2012tidal}.

We gather that the peak frequencies are weakly dependent on $\eta$ while the amplitudes of GW polarisation states are proportional to $\eta$ and therefore maximum for  equal mass compact binaries. 
Therefore, our BWM approach will be more suitable to constrain comparable mass hyperbolic compact binaries with total mass less than $30\,M_{\odot}$ in the LVK datasets as their peak frequencies lie above 30 Hz.
However, it is possible that higher total mass events are still possible LVK sources as such events can provide higher harmonics with substantial power that fall in the LVK-sensitive frequency window due to the narrow-band nature of such signals.
Influenced by these considerations, we now provide details of the approximant
that is used here to compute the distance reach of hyperbolic events and which could be helpful to search for such transient GW events in the data streams of GW observatories.

\begin{figure*}
\includegraphics[width=0.85\linewidth]{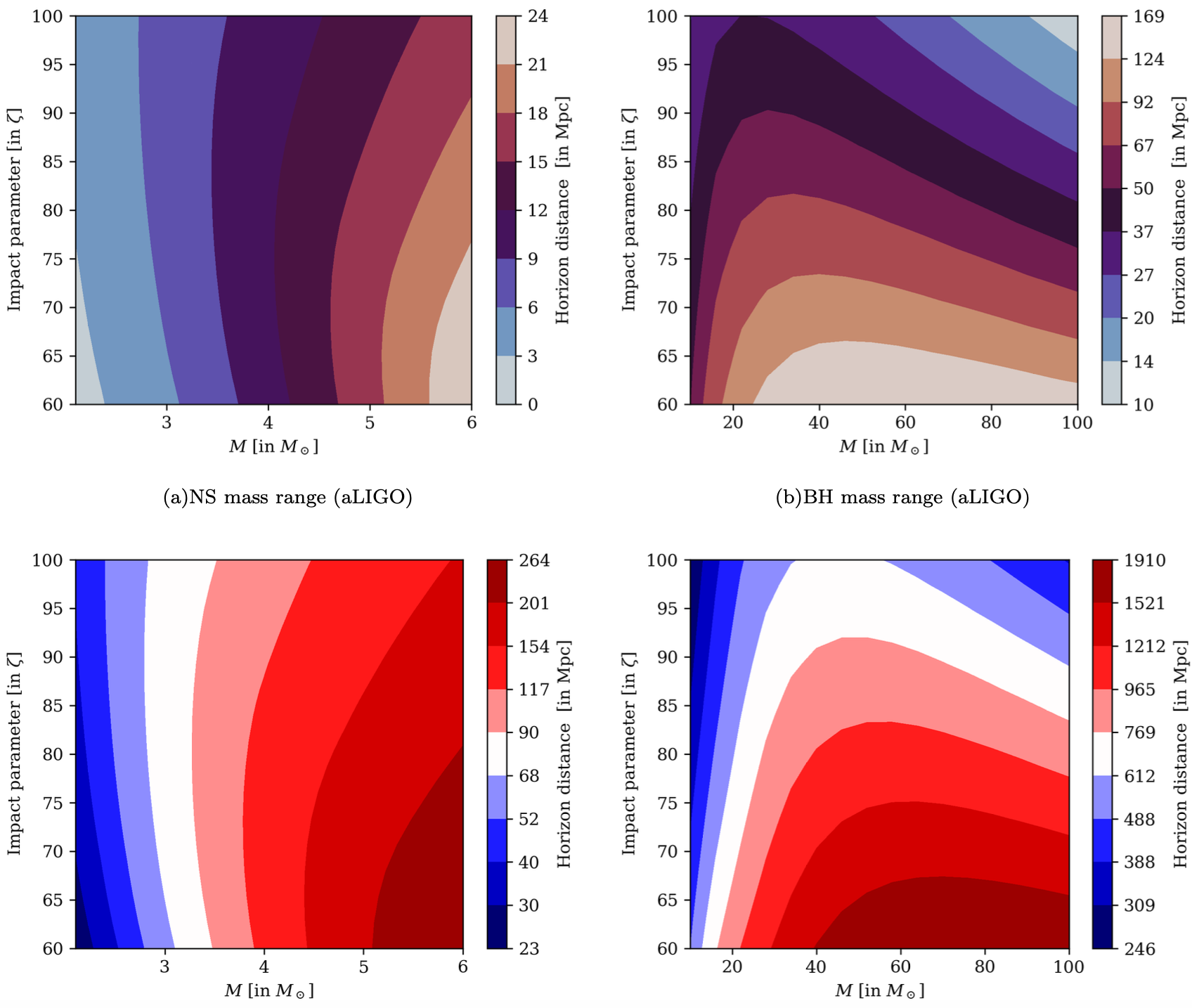}
\caption{
Horizon distance contour plots of hyperbolic encounters with SNR$=8$ for aLIGO and ET observatories while imposing the restriction that $r_\text{min} \gtrsim 10  \, \zeta $. 
We let compact binaries have equal mass and choose $e=1.15$.  
The median reach of NS events for aLIGO is $\sim 15$ Mpc and it is around $80$ Mpc for BH events.
We infer that BH systems with $M$ in the $30-80\, M_{\odot}$ range with $b$ around $60\, \zeta$ should be 
visible up to $170$ Mpc. This consideration prompted us to explore the aLIGO distance reach of hyperbolic events while relaxing the above $r_\text{min}$ restriction in Fig.~\ref{horizon-cons_e-all-zone}.
}
\label{horizon-heatmap-conse}
\end{figure*}

\subsection{Detectability of transient hyperbolic events with ground-based GW observatories}

We begin by presenting the details of our PN-accurate waveform family 
called \texttt{HyperbolicTD}.
These routines for waveform generation from hyperbolic encounters are compatible with \texttt{LAL} C99, the standard code basis of the LIGO Algorithm Library Suite (LALSuite), and implemented in LALSimulation, the package of routines for waveform and noise generation in LALSuite~\citep{lalsuite_official}. 
Given an initial eccentricity, impact parameter, and eccentric anomaly, the code first converts the impact parameter to the PN-parameter $x$ using the PN-accurate inversion of Eq.~\eqref{b_hyp}. 
Thereafter, it evolves a system of three coupled differential equations, namely  $\frac{d x}{dt}$ and $\frac{d e_t}{dt}$, given by Eqs.~\eqref{gw_em}, and a 3PN-accurate expression for $\frac{d u}{dt}$ provided by Eq.~\eqref{dudt}. 
It should be noted that $\frac{d u}{dt}$ begins at the Newtonian order, whereas the evolution equations for $x$ and $e_t$ start only from the 2.5 PN (absolute) order. 
The resulting temporal evolution for $u$, $x$ and $e_t$ are incorporated into the 3PN-accurate expressions for the dynamical variables, namely $r$, $\dot r$, $\phi$ and $\dot \phi$.
In the final step, these temporally varying dynamical variables are imposed  in the quadrupolar order GW polarization states, given in Eq.~\eqref{hcpQ}. 
Further, we let $r_\text{min}$ remain above $10\, \zeta$ to ensure the validity of PN approximation for describing these events.

Henceforth, the waveforms are called by specifying the masses of the binary, the impact parameter, and the eccentricity. Additionally, extrinsic parameters like the inclination angle, a reference phase $\phi_0$, and the distance to the source can be chosen.
If a minimum frequency is specified, the waveform is high-pass filtered.
To avoid artifacts at the beginning and the end of our templates, it is tapered such that the waveform consistently starts at zero and ends at zero amplitude. 
Clearly, this procedure gets rid of the linear memory that might be present. 
However, all frequencies above the minimum frequency are still accurately represented.

Using our \texttt{HyperbolicTD} approximant, we have computed horizon distances of 
the ground-based second and third generation GW detectors for hyperbolic encounters.
Recall that this distance refers to the farthest luminosity distance a given source could ever be detected above the threshold at an optimal sky location and binary inclination/orientation.
For the present investigation, we let the detection threshold be a matched-filter signal-to-noise ratio (SNR) of 8.  We note that though this horizon distance is a measure of the furthest reach of a GW observatory, it is not representative of the general population as the detector response patterns are not spherical.
We have chosen aLIGO and ET as typical representatives for these observatories. 

Following the general practice, the matched-filter SNR of a template $g(t)$, given a time series $h(t)$, is computed by
\begin{align}
    \mathrm{SNR} = \frac{\langle h, g \rangle}{\sqrt{\langle g, g \rangle}} \,,
\end{align}
and the inner product is defined as
\begin{align}
    \langle a, b \rangle = 4 \mathrm{Re} \int_0^\infty \frac{\tilde a(f) \tilde b^* (f)}{S_h (f)} df \,,
\end{align}
where $S_h (f)$ is the noise power spectral density (PSD) of the detector.
For the $(M,b)$ parameter space for which we explored the peak frequencies in Fig.~\ref{peakfreq-heatmap}, we compute the horizon distance, i.e., the distance at which the corresponding waveform template is found with a matched-filter SNR of 8., using the public code provided in Ref.~\cite{chen2017distances} which accurately takes into account cosmological effects. Recall that due to the expansion of the universe, the GW signal is red-shifted, and therefore the GW signal from a binary with total mass $M$ at redshift $z$ appears to have a total mass of $(1+z) M$ when observed on earth. 
The horizon distances are displayed in Fig.~\ref{horizon-heatmap-conse} for both NS and BH binaries while keeping $e_t = 1.15$. For this particular study we have employed the zero detuned Advanced LIGO noise PSD~\cite{LIGO2018aLIGOPSD} as a representative for the current generation of ground based detectors (LIGO, Virgo, KAGRA~\cite{t2021overview}) and the ET-D noise PSD~\cite{hild2011ETPSD} as a representative of the proposed third generation of ground based detectors (Einstein Telescope, Cosmic Explorer~\cite{hall2022cosmic}).  
Note that the total masses quoted are the source frame masses and that the horizon distance should be interpreted as a luminosity distance.    

From the plots given in Fig.~\ref{horizon-heatmap-conse}, we infer that optimally placed NS and BH binaries are visible for aLIGO up to $\sim$ 20 and 170 Mpc, respectively when we impose the restriction that $r_\text{min} \gtrsim 10 \, \zeta$, their counterparts for ET being $\sim 260$ Mpc and $\sim 1.9$ Gpc, respectively.
Further, typical NS binary events are only visible to 10 Mpc which makes such events highly unlikely to be detected in the era of second-generation GW observatories.
We observe that the distance reach for BH binary events decreases as we increase the impact parameter; this is due to the fact that peak frequencies move out of the sensitive frequency window of these observatories and additionally, waveform amplitudes decrease as we increase impact parameters.
A similar explanation holds for the observation that the distance reach approaches a peak and then decreases when we increase the total mass for a given value of $b$. 
These considerations prompted us to relax the restriction that $r_\text{min}= 10\, \zeta$ and explore its consequences in Appendix~\ref{AppA}.

The plots in Appendix~\ref{AppA} reveal that 
aLIGO will be able to observe hyperbolic events up to $\sim 500$ Mpc distances for equal mass BH binaries having  $M$ in the $50-80\,M_{\odot}$ range with 
$b \sim 50 \, \zeta $ and $e_t \sim 1.1$. 
This is a promising inference, provided our PN approximation works for these hyperbolic configurations.
In the next section, we explore the PTA implications of our PN-accurate description for  hyperbolic encounters.

\section{Modeling PTA responses to  Hyperbolic Events}
\label{Sec3}
It should be obvious by now that hyperbolic encounters of two BHs lead to GW burst signals.
An important feature of the resulting gravitational waveform is the presence of certain linear GW memory \cite{devittori2014gravitational}.
An appropriate way to demonstrate the presence of such an effect is to take the $t \rightarrow +\infty$ limit of $h_{\times|Q}$ given by Eq.~\ref{hcpQ} while using Newtonian-accurate expressions for various dynamical variables.
This leads to
\begin{align}
h_{\times|Q} &=
-\frac{2\, G\, M\,\eta}{c^4\, R'} \, v_{\infty}^2 \, \sin 2\, \phi_{\infty}
\,,
\end{align}
where $\phi_{\infty}$ stands for the orbital phase at $\pm \infty$ while $v_{\infty}$ provides the value of $\dot r(u) $ at $ t \rightarrow \pm \infty$ (it turns out that all other dynamical variables   $G\,M/r$ and $\dot \phi$ vanish at these limits).
The fact that the above expression is an odd function of $\phi$ implies that
\begin{subequations}
\begin{align}
\lim_{x \to +\infty} h_{\times|Q} &= -\lim_{x \to -\infty} h_{\times|Q} \,, \\
\delta h_{\times} &= 
-\frac{4\, G\, M\,\eta}{c^4\, R'} \, v_{\infty}^2 \, \sin 2\, \phi_{\infty}\,,
\end{align}
\end{subequations}
where $\delta h_{\times} = \lim_{x \to +\infty} h_{\times|Q} - \lim_{x \to -\infty} h_{\times|Q}$.
This essentially explains the presence of the linear GW memory in hyperbolic passages of SMBHs, which is visible as the non-zero offset in the $ h_{\times}$ plot in Fig.~\ref{allptapsi}.
It is usual to term such a constant non-zero $h_{\times}$  offset  as certain non-oscillatory GW effects associated with such burst signals \cite{braginskii1987gwb}.
We note that Ref.~\cite{finn2010detection} provided a detailed prescription to detect and characterize GW burst signals, possibly from SMBHs in Newtonian parabolic orbits by PTAs.
Further, there are ongoing efforts to build algorithms to search for GW burst events in PTA datasets \cite{wang2015searching,aggarwal2020nanograv,becsy2021bayesian}.
These considerations prompted us to  explore the PTA implications of our GW burst signals and their associated linear memory effect.

There are multiple ongoing efforts to constrain non-linear GW memory events in the various PTA datasets and these events are usually associated with GWs from SMBH binary coalescence  \cite[e.g.][]{wang2015searching,aggarwal2020memory}.
The non-linear GW memory arises due to the fact that certain hereditary contributions to  GWs from BH binary coalescence  themselves follow unbound trajectories \cite{blanchet1992hereditary,favata2011memory}.
Therefore, such GW memory signals should allow us to set limits on the rate of SMBH binary coalescence events. 
For example, a recent NANOGrav effort  provided a limit on the rate of non-linear GW memory events to be below $0.4$/yr and the associated strain puts a similar constraint on SMBH binary coalescence with certain optimum BH masses and orbital inclinations up to 1 Gpc \cite{aggarwal2020memory}.
We note that these events are detectable via relatively sudden changes in the apparent pulse frequency of the PTA pulsars, and the sensitivity of PTA experiments to  GW memory events was discussed in Refs.~\cite{vanstraten2010gravitational,cordes2012detecting}. 
Further, Ref.~\cite{madison2017pulsar} derived the PTA responses to burst with memory events originating from near-field sources such as supernovae and compact binary mergers.
Interestingly, these memory events are not restricted to BH binaries as it turns out that the tidal effects associated with compact binaries involving NSs can also be captured by the underlying nonlinear GW memory effect and are therefore  relevant to terrestrial GW observatories \cite{tiwari2021leveraging}.
We also note in passing that a similar application of the GW phasing approach for computing the PTA signals induced by supermassive eccentric binaries was pursued in Refs. \cite{susobhanan2020pulsar,susobhanan2022cost}.


In what follows, we provide the details of our ready-to-use package that should be useful to constrain  hyperbolic passages of SMBHs using their inherent linear GW memory effect.
We begin by describing briefly our approach to obtain the timing residuals that are induced by  SMBH binaries in 3PN-accurate hyperbolic orbits.

\subsection{PTA signals associated with our BWM GW Events }
\label{sec:pta-signal}

A GW passing across the line of sight of a pulsar will induce temporally evolving modulations in the pulsar's observed times of arrival (TOAs).
These modulations, termed the GW-induced (pre-fit) timing residuals or the PTA signal, are given by \cite{anholm2009optimal}
\begin{equation}
    R(t_E) = \int_{t_0}^{t_E} (h(t_E') - h(t_E'-\Delta_p)) dt_E'\,,
\end{equation}
where $t_0$ is an arbitrary fiducial time and $\Delta_p$ stands for the geometrical time delay given by $\Delta_p=D_p(1-\cos\mu)/c$, where $D_p$ is the distance to the pulsar, and $\mu$ is the angle between the line of sight to the pulsar and the GW source.
Further, the time variables $t_E$ and $t_E'$ are usually measured in the solar system barycenter (SSB) frame while $t_E$ relates to the typical coordinate time measured in the GW source frame via the cosmological redshift
\begin{equation}
    t_E-t_0 = (1+z)(t-t_0)\,.
    \label{eq:cosmo-z}
\end{equation}
The temporally evolving dimensionless GW strain $h$ is given by
\begin{equation}
h=\begin{bmatrix}F_{+} & F_{\times}\end{bmatrix}\begin{bmatrix}\cos2\psi & -\sin2\psi\\
\sin2\psi & \cos2\psi
\end{bmatrix}\begin{bmatrix}h_{+}\\
h_{\times}
\end{bmatrix}\,,
\label{eq:h_pta}
\end{equation}
where 
$F_{+,\times}$ are the antenna pattern functions that depend on the sky locations of the GW source and the pulsar, and $\psi$ is the usual GW polarization angle.
The explicit expressions for $F_{+,\times}$ in terms of the sky coordinates of the pulsar and the GW source may be found in, e.g.,  Ref.~\cite{lee2011gravitational}.

It is convenient to define the following two quantities
\begin{equation}
    s_{+,\times}(t_E) = \int_{t_0}^{t_E} h_{+,\times}(t_E') dt_E'
    =(1+z)\int_{t_0}^t h_{+,\times}(t') dt'\,,
\end{equation}
such that we can express $R(t_E)$ as
\begin{align}
    R(t_E) &= \begin{bmatrix}F_{+} & F_{\times}\end{bmatrix}\begin{bmatrix}\cos2\psi & -\sin2\psi\\
\sin2\psi & \cos2\psi
\end{bmatrix}\nonumber\\
&\qquad \times \begin{bmatrix}s_{+}(t_E)-s_{+}(t_E-\Delta_p)\\
s_{\times}(t_E)-s_{\times}(t_E-\Delta_p)
\end{bmatrix}\,.
\end{align}
In the case of transient GW events such as  our hyperbolic  encounters of BHs, 
we may drop the pulsar terms, namely $s_{+,\times}(t_E-\Delta_{p})$ contributions
in the above expression.
This is because the PTA pulsars are typically hundreds to thousands of light years away from the SSB and therefore  the duration of the transient will be much shorter than $\Delta_p$.
This ensures that $s_{+,\times}(t_E-\Delta_{p})$ terms produce no measurable contributions to the PTA signals. 
Therefore, for our transient events, we write without any loss of generality 
\begin{align}
    R(t_E) &= \begin{bmatrix}F_{+} & F_{\times}\end{bmatrix}\begin{bmatrix}\cos2\psi & -\sin2\psi\\
\sin2\psi & \cos2\psi
\end{bmatrix}\begin{bmatrix}s_{+}(t_E)\\
s_{\times}(t_E)
\end{bmatrix}\,.
\label{eq:R(t_E)_transient}
\end{align}



For the present effort, we employ the quadrupolar order expressions for $h_{\times,+}$, given by Eqs.~\eqref{hcpQ}. 
Further, we employ 3PN-accurate expressions for $r$, $\dot r$, $\dot \phi$, and $\phi$ in terms of $n$, $e_t$, and $u$ (in other words, 3PN extensions of Eqs.~\eqref{orbit_1}).
To incorporate temporal evolution of $u$, we employ the following 3PN-accurate Kepler Equation, adapted from Ref.~\cite{cho2018gravitational}
\begin{widetext}
\begin{align}
\label{eq:ltou}
l= n\, (t- t_0) &=(e_t \sinh u-u)-\frac{x^2 (12 \nu (-5+2 \eta )+e_t (-15+\eta ) \eta  \sin(\nu))}{8 \sqrt{e_t^2-1}}+\frac{x^3}{6720 (e_t^2-1)^{3/2}}
\bigg{\{}e_t (67200\notag \\[2ex]
&-3 (-47956+105 e_t^2+1435 \pi ^2) \eta -105 (592+135 e_t^2) \eta ^2+35 (-8+65 e_t^2)
\eta ^3) \sin \nu\notag \\[2ex]
&+35 ( (8640-13184 \eta +123 \pi ^2 \eta +960 \eta ^2+96 e_t^2 (30-29 \eta +11 \eta ^2)) \, \nu+12
e_t^2 \eta  (116-49 \eta +3 \eta ^2) \notag \\[2ex]
& \times \sin(2 \nu)+e_t^3 \, \eta  (23-73 \eta +13 \eta ^2) \sin(3 \nu))\bigg{\}} \ ,\end{align}   
\end{widetext}
and  we  are required to express  $ \sqrt{\frac{e_\phi+1}{e_\phi-1}} $ in terms of $e_t$ and $x$ using Eq.~\eqref{eq:ephiet}.
To tackle the above transcendental equation,  we introduce a new `auxiliary eccentric anomaly' influenced by Ref.~\cite{cho2022generalized}.
The idea is to introduce $ \hat u$ such that the above equation takes the form of the classical Kepler Equation, namely 
\begin{equation}
 l= \hat{u}- e_t\, \sinh \hat u\,.
\label{eq:kepler}
\end{equation}
The main motivation for introducing $\hat u$ is to employ Mikkola’s method, detailed in Ref.~\cite{mikkola1987cubic}, to solve the $\hat u$ classical Kepler Equation.
We may recall that the 
Mikkola’s method is an optimized and highly accurate numerical approach to
tackle the classical Kepler equation \cite{colwell1993solving}.
This procedure provides an accurate and efficient way to obtain $\hat u$ as a function of the coordinate time or  $l$. 
Thereafter, we employ the following PN-accurate expression that provides  $u$ in terms of $\hat u$ 
 \begin{widetext}
     \begin{align}
     \label{uinuhat}
	u&=\hat{u} +\frac{x^2}{8 \, (e_t \cosh \hat{u}-1)^2} \bigg{\{}\frac{24 (-5+2 \eta ) \, (e_t \cosh \hat{u}-1)}{\sqrt{e_t^2-1}} \, \arctan \left[\sqrt{\frac{e_t+1}{e_t-1}} \, \tanh\left(\frac{\hat{u} }{2}\right)\right] +e_t \, (\eta-15 ) \eta  \sinh\, \hat{u}\bigg{\}}\notag \\[2ex]
	&+x^3 \bigg{\{}\frac{1}{8
		(e_t^2-1) (e_t \cosh \hat{u}-1)^3} \, \bigg{[}e_t
	(-4+\eta ) (-60+3 \, (8+5 e_t^2) \eta -e_t^2 \eta ^2+e_t (60-39 \eta +\eta ^2) \cosh\, \hat{u}) \sinh\, \hat{u}\bigg{]}\notag \\[2ex]
	&-\frac{1}{6720 (e_t^2-1)^{3/2} (e_t \cosh \hat{u}-1)} \, \Bigg{(}\frac{1}{e_t \cosh u-1}\bigg{[}e_t
	\sqrt{e_t^2-1} (67200-3 (-47956+105 e_t^2+1435 \pi ^2) \eta \notag \\[2ex]
	&-105 (592+135 e_t^2) \eta ^2+35 (-8+65
	e_t^2) \eta ^3) \sinh\, \hat{u}\bigg{]}+35 \bigg{(} \bigg{[}8640+(-13184+123 \pi ^2) \eta +960 \eta
	^2+96 e_t^2 (30-29 \eta \notag \\[2ex]
	&+11 \eta ^2)\bigg{]} \, 2\, \arctan \left(\sqrt{\frac{e_t+1}{e_t-1}} \,  \tanh \frac{\hat{u} }{2}\right)+\frac{1}{(e_t \cosh \hat{u}-1)^2}\bigg{[}24 \, 
	e_t^2 \, \sqrt{e_t^2-1} \,  \eta  \, (116-49 \eta +3 \eta ^2) \, (e_t-\cosh\, \hat{u}) \sinh\, \hat{u}\bigg{]}\notag \\[2ex]
	& -\frac{1}{2\, (e_t \cosh \hat{u}-1)^3}  \bigg{[}e_t^3 \, 
	\sqrt{e_t^2-1}  \, \eta \,  (23-73 \eta +13 \eta ^2) \, (-2-7 e_t^2 +12 e_t \cosh\, \hat{u}+(-4+e_t^2)\,  \cosh 2 \hat{u}) \, \sinh\, \hat{u} \bigg{]}\bigg{)}\Bigg{)}\bigg{\}}.
\end{align}
 \end{widetext}
 

With the help of resulting $u$ values and our 3PN-accurate versions of Eqs.~(\ref{orbit_1}), we obtain temporally evolving quadrupolar order $h_{\times,+}$ due to SMBH binaries in  3PN-accurate hyperbolic orbits.
Thereafter, we incorporate the effects of GW emission on the above 3PN-accurate conservative dynamics by solving three coupled PN-accurate differential equations, namely $dx/dt$, $d e_t/dt$ expressions, given by Eqs.~\eqref{gw_em} and $dl/dt=n$, and we employ the expressions that are in the modified harmonic gauge.
This naturally leads to temporally evolving quadrupolar order $h_{+,\times}$ due to SMBH binaries in fully 3PN-accurate hyperbolic orbits.

It is now straightforward to obtain ready-to-use PTA responses to  our temporally evolving  $h_{\times,+|Q}$ due to SMBH binaries in fully 3PN-accurate hyperbolic orbits.
It should be evident that it will not be possible to perform analytically the integrals that appear in the expression for $R(t)$, given by Eq.~\eqref{eq:R(t_E)_transient}.  
Therefore, we perform these integrations numerically, similar to what is pursued in Ref.~\cite{susobhanan2020pulsar}, where numerical integration is applied to compute the PTA signals due to relativistic eccentric binary systems.
In what follows, we describe various facets of our \texttt{Python}  package \texttt{GW\_hyp} for computing the PTA signals, as detailed in this section, that should be useful to search for such GW events in PTA datasets. 



\subsection{Details of the \texttt{GW\_hyp} package and its deliverables }
\label{sec:GW_hyp}


 We provide in this subsection a schematic description of a \texttt{Python} package   \texttt{GW\_hyp}\footnote{ \url{https://github.com/subhajitphy/GW_hyp/}} that  implements the PTA signal  described in the previous 
 subsection.
 This software package follows the following steps to 
compute $R(t_E)$, given a set of TOAs.
\begin{enumerate}
    \item Convert the TOAs to the source frame by applying the cosmological redshift (Eq.~\ref{eq:cosmo-z}).
    \item Evaluate $n$, $e$, and $l$ at a dense uniform sample of times that span the TOA range in the source frame by numerically integrating the reactive evolution equations (Eqs.~\ref{gw_em})
    along with the $dl/dt =n$ equation.
    This is done using the \texttt{scipy.integrate.odeint} function.
    \item Solve the Kepler equation by 
    invoking Mikkola's method 
    with the help of Eqs.~(\ref{eq:kepler}) and (\ref{uinuhat}) 
    to obtain the eccentric anomaly $u$ for each value of $l,n$ and $e$.
    \item Compute $r(u)$, $\dot{r}(u)$, $\phi(u)$, and $\dot{\phi}(u)$ 
 [Eqs.~\eqref{orbit_1}].
    \item Compute $h_{+,\times}(r,\dot{r},\phi,\dot{\phi})$ by employing Eqs.~(\ref{hcpQ}).
    \item Compute $h(t)$ [Eq. \eqref{eq:h_pta}].
    \item Integrate $h(t)$ to obtain $R(t)$ using the \texttt{scipy.integrate.cumtrapz} function that implements the trapezoidal method.
    \item Evaluate the PTA signal at each TOA by interpolating the dense $R(t)$ samples. This is done using the \texttt{scipy.interpolate.CubicSpline} class.
\end{enumerate}

We have implemented a top-level function named \texttt{GW\_hyp.hyp\_pta\_res} that produces the PTA signals given an \texttt{ENTERPRISE} pulsar object and a set of source parameters.
This function can readily be invoked to create an \texttt{ENTERPRISE} `\texttt{Signal}' object to search for GWs from hyperbolic encounters of SMBHs in PTA datasets.
\par

\subsection{Pictorial exploration of \texorpdfstring{$R(t)$}{R(t)} due to scattering BH systems in relativistic hyperbolic orbit}

\begin{figure*}
\centering
\includegraphics[width=0.8\linewidth]{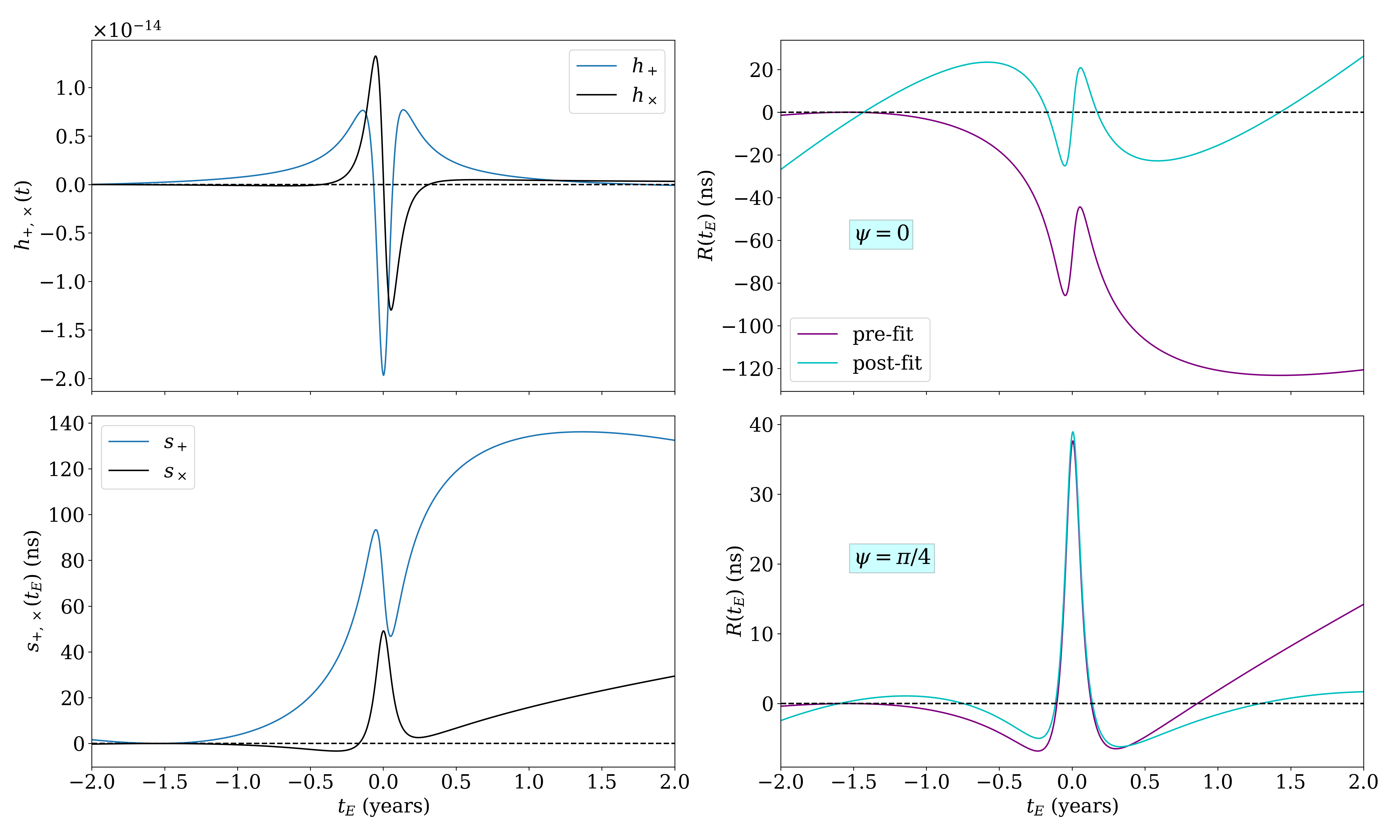}
\caption{Plots for 
the quadrupolar order 
$h_{+,\times}(t), \ s_{+,\times}(t_E)$, pre-fit and post-fit residuals induced on PSR 1909-3744 
by an SMBH binary scattering event, characterized by 
$M=10^{10} M_\odot, \ b=70 \, \zeta , e_t=1.1$ and $i=\pi/3$. 
We let the polarization angle($\psi$)  to take two values ($\psi=0$ and $\psi=\pi/4$) 
and the event is occurring at a  red-shift $z=0.3$ ($R^\prime \sim 1.6$ Gpc).
The choice of $M$ is influenced by the SMBH in M87 while $e_t$ and $b$ values are chosen 
so that the peak GW frequency is around $\sim 10^{-8}$Hz.
Interestingly, the shape and magnitudes of these timing residuals depend on $\psi$.}
\label{allptapsi}
\end{figure*}

\begin{figure*}
\centering
\includegraphics[width=0.85\linewidth]{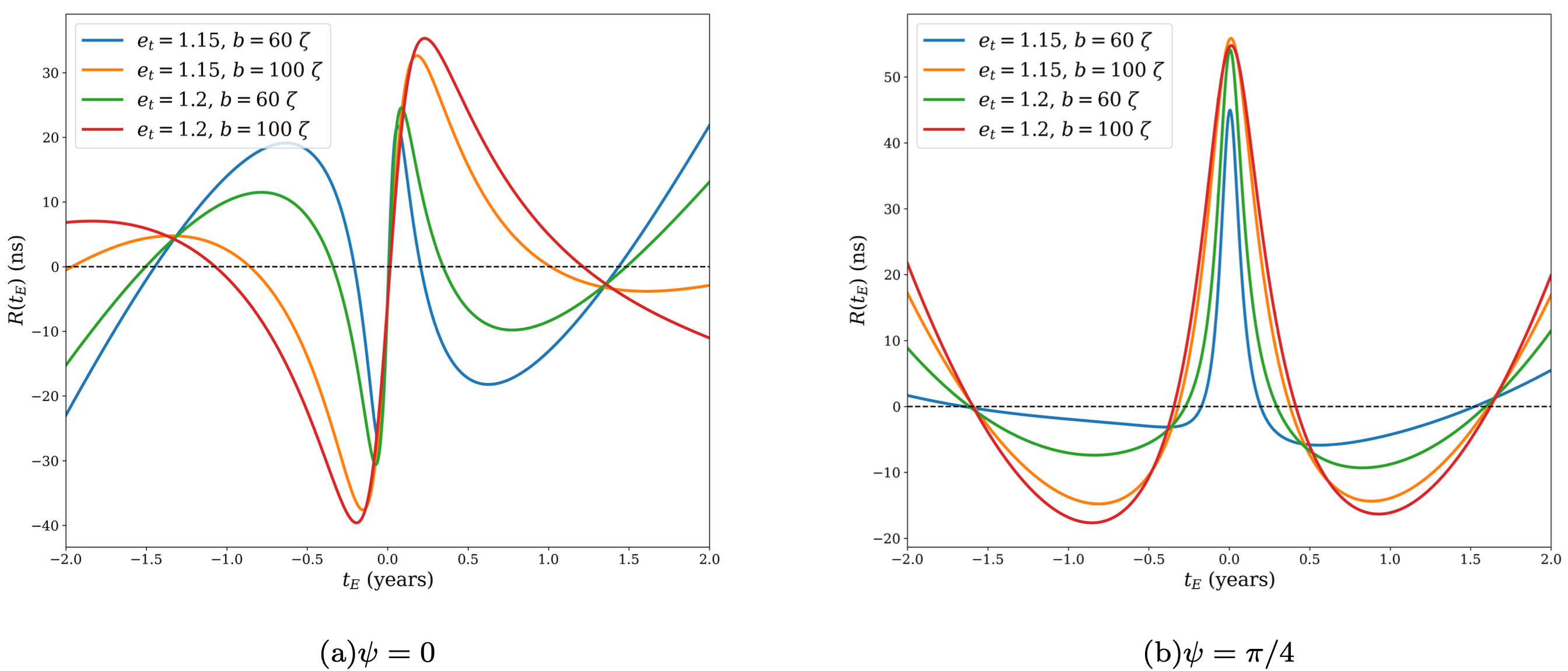}
\caption{Plots for the 
post-fit timing residuals while allowing eccentricities and impact parameters to take several values 
for two different polarization angles $(\psi)$ of the GW source.
All other parameters are kept the same as in Fig~ \ref{allptapsi}  and the peak GW frequencies 
lie between $3 \times 10^{-8}$ and $9\times10^{-8}$ Hz. }
\label{fig:varyeb_pta}
\end{figure*}

We begin by displaying various aspects of the PTA responses to GWs from hyperbolic encounters while considering a fiducial equal mass $(q=1)$ BH binary system with $M=10^{10} \, M_\odot$ at a luminosity distance of 1.6 Gpc ($z=0.3$) with an  inclination angle of $i=\pi/3$  in Fig.~\ref{allptapsi}.
We focus on  PSR 1909-3744  and let the eccentricity and the impact parameter be fixed at $e_t=1.1$ and $b=70 \, M$, respectively. 
Further, we choose the sky location of the GW source to be RA $19^h00^m00^s$, DEC $0^{\circ}$ while allowing two values for the polarization angle ($\psi=0$ and $\pi/4$).
We present certain post-fit residuals that are acquired after eliminating the constant, linear, and quadratic terms from GW-induced pre-fit timing residual, namely $R(t_E)$. 
In particular, we fit a quadratic function of the form $f(t)=a_0+a_1 \, t+a_2 \, t^2$, to the GW-induced pre-fit residual $R(t)$  and subtract $f(t)$ from the pre-fit residual to obtain post-fit residual.
We infer from the plots in Fig.~\ref{allptapsi} that measurable timing residuals depend on the $\psi$ values. 
To probe the dependencies on post-fit residuals on $b$ and $e_t$ values, we plot Fig.~\ref{fig:varyeb_pta} while keeping all other parameters as in Fig.~\ref{allptapsi}.

We now proceed to display the quadrupolar nature of our PTA signal in Figs.~\ref{fig:skyplotpsi0}, and \ref{fig:skyplotpsi45}.
The heat maps  in these figures show responses of pulsars, distributed across the sky, to a given GW event and these 
responses are  essentially 
the differences between the  maximum and the minimum of post-fit $R(t_E)$ within a given time span as described in Ref.~\cite{susobhanan2020pulsar}. 
Note that 
Fig.~\ref{fig:skyplotpsi0}(a) and Fig.~\ref{fig:skyplotpsi45}(a) show essentially post-fit $R(t_E)$ strengths for $\psi=0$ and $\psi=45^\circ$ values, respectively.
Additionally, we display 
the post-fit residuals associated with the Newtonian and 3PN accurate hyperbolic orbits for the well-known PSR J1909-3744. 
This pulsar is 
chosen as it gives one of the best strengths of $R(t_E)$ which should be evident from the sky sensitivity plot.
For these plots, we let the impact parameter of the scattering system  fixed at $b=70 \ \zeta $ and all other parameters are the same as in Fig.~\ref{allptapsi}. Interestingly, PN corrections can reduce the magnitudes of $R(t_E)$ as well as its shape.  
The fact that these cosmological events induce 
timing residuals in the range of 50 nanoseconds suggest that the present and upcoming IPTA data releases should be able to 
provide interesting astrophysical constraints on the occurrence rate of such events.

\begin{widetext}

\begin{figure*}
\centering  
\includegraphics[width=0.85\linewidth]{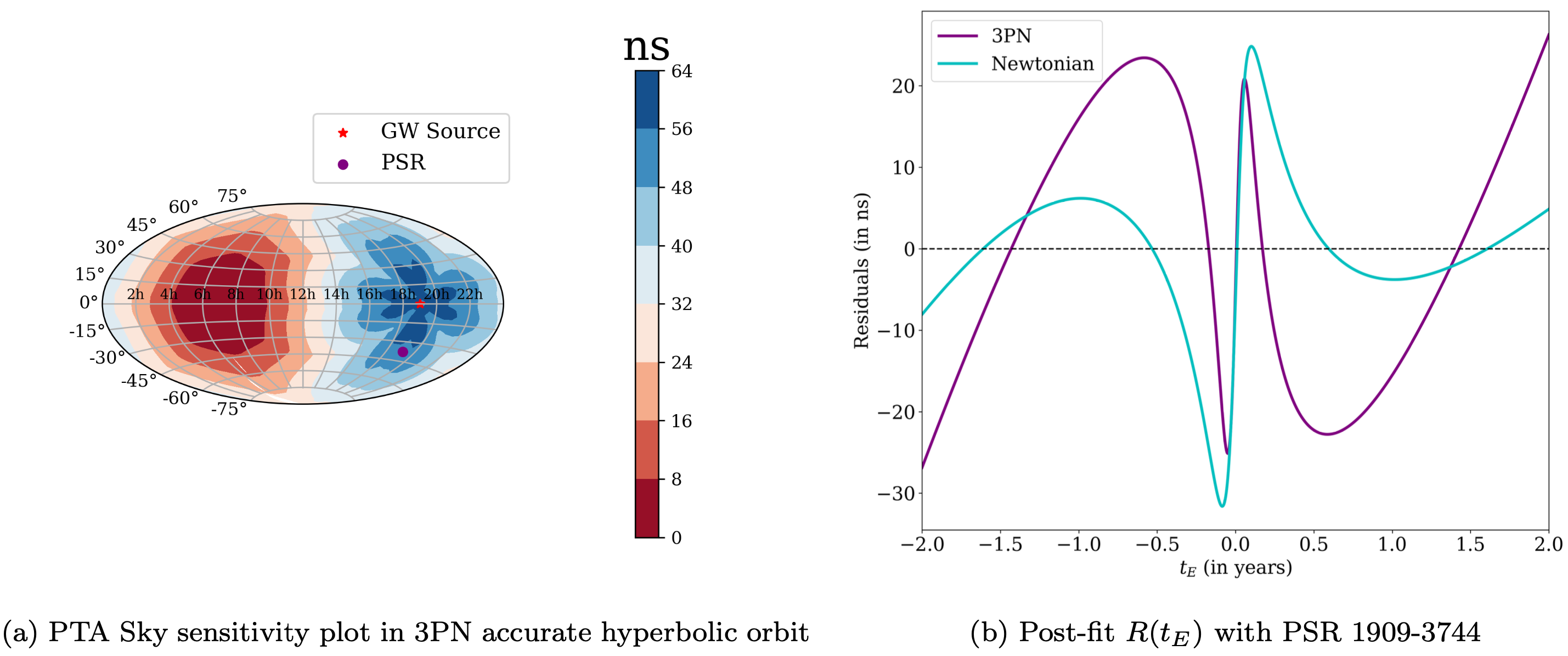}
\caption{
Heat map of the post-fit $R(t_E)$ strengths for pulsars distributed across the sky where the red
and purple dots stand for the  sky locations of the GW source and PSR 1909-3744, respectively.
 Additionally, we plot the actual post-fit $R(t_E)$ while 
considering Newtonian and PN-accurate orbital description for scattering event that is characterized by 
$M= 10^{10} M_\odot$, $z=0.3 \ (R^\prime=1.6$ Gpc), $e_t=1.1$, $b=70 \, \zeta$, $f_\text{peak}=9.43 \times 10^{-8}$ Hz and $\psi=0^\circ$.}

\label{fig:skyplotpsi0}
\end{figure*}

 \begin{figure*}
\centering  
\includegraphics[width=0.85\linewidth]{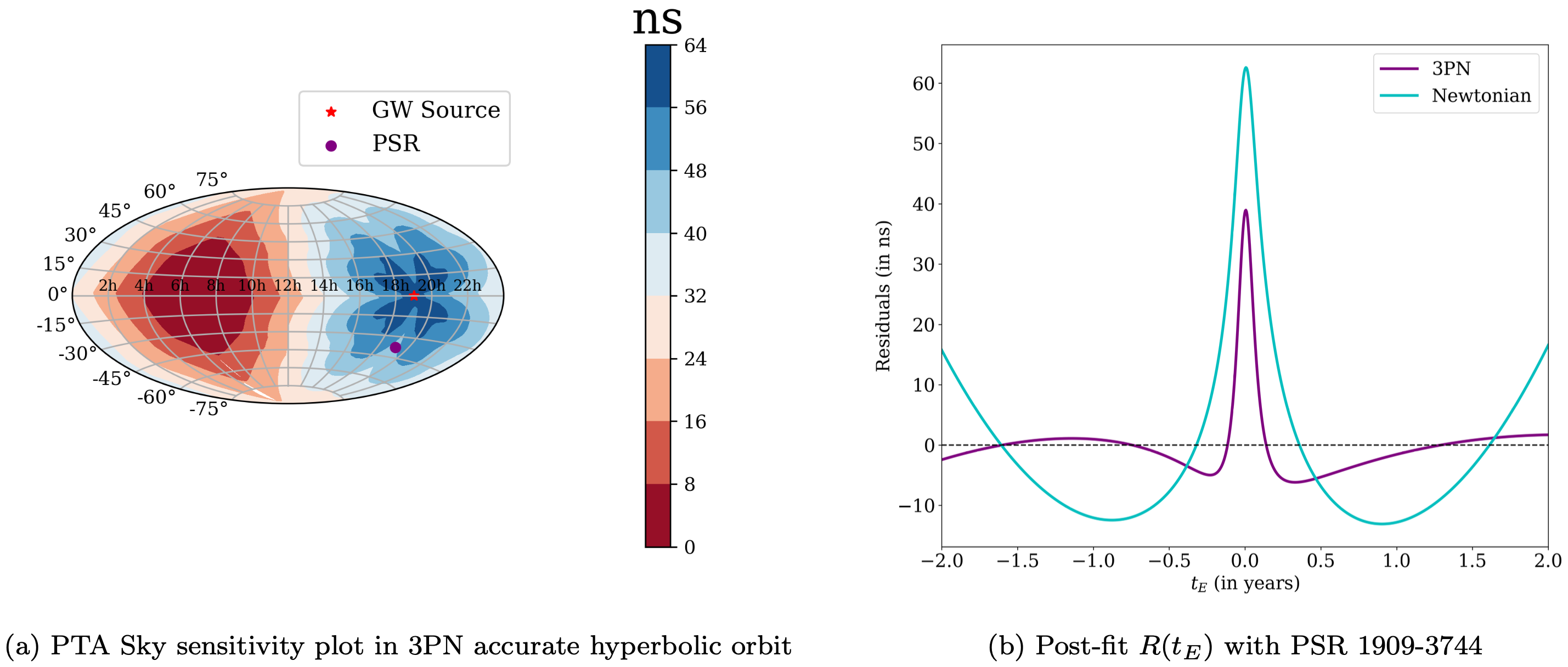}
\caption{
Plots similar to those displayed in Fig.~\ref{fig:skyplotpsi0} while allowing 
$\psi=45^\circ$. We clearly see the influence of $\psi$ on the post-fit residuals.
}
\label{fig:skyplotpsi45}
\end{figure*}
\end{widetext}

\section{Conclusions}

We provided detailed descriptions of two waveform packages, namely \texttt{LAL}-compatible \texttt{HyperbolicTD} and \texttt{ENTERPRISE}-compatible \texttt{GW\_hyp}, that model  GWs from hyperbolic encounters between comparable mass non-spinning  compact objects. 
These packages employ 3PN-accurate Keplerian type parametric solution, detailed in Ref.~\cite{cho2018gravitational}, and GW phasing  approach of Ref.~\cite{damour2004phasing} to describe general relativistic  trajectories of compact objects in unbound orbits.
Our \texttt{HyperbolicTD} approximant will be useful to search and characterize such transient burst signals in the datasets of ground-based GW observatories.
We note that although \texttt{HyperbolicTD} is already implemented in a way that is fully compatible with the GW data analysis routines in the LIGO Algorithm Library Suite, we plan to submit the waveform model to an internal review by the LIGO-Virgo-KAGRA collaborations before publishing it in an official LALSuite release.
We pursued preliminary data analysis investigations with aLIGO at design sensitivity to conclude that hyperbolic events involving stellar mass BH binaries are potential GW sources for the current generation of ground-based instruments.
The present investigations reveal that hyperbolic encounters that involve BHs having $M \sim 80 \, M_\odot$ with $e_t \sim 1.1$ and $ b\sim 50 \, \zeta$ should be visible to 500~Mpc distances by aLIGO.
Investigating GW prospects for galaxy-based observatories, we find that  our \texttt{ENTERPRISE} compatible \texttt{GW\_hyp} package should be relevant to the search for BWM events in the PTA datasets.
Our post-fit timing residual plots reveal that hyperbolic encounters involving $10^{9}\, M_{\odot}$ SMBHs 
should be detectable by the SKA era PTA at cosmological distances. Interestingly, PN approximation should be
valid to describe such events if their peak frequencies are in the nano-hertz range.

To further explore the observational prospects of hyberbolic encounters for the current generation of ground-based detectors, we are probing possibilities for source characterisation after a successful detection of such events. We are pursuing a detailed parameter estimation study with our \texttt{HyperbolicTD} approximant, using a Bayesian inference library for gravitational-wave astronomy, \textsc{Bilby}~\cite{ashton2019bilby}, to recover the source characteristics of fiducial synthetic GW signals from hyperbolic encounters.
However, a detailed comparison of GW polarization states from the 
\texttt{HyperbolicTD} approximant with their numerical relativity
counterparts has not yet been performed, due to a lack of available NR waveforms, and will be crucial to validate the use of PN approximation to describe a broad parameter space of hyperbolic encounters. 
Moreover, it will be interesting to employ hyperbolic $h(t)$ that arise from the Effective One Body approach, detailed in Ref.~\cite{nagar2021effective,gamba2022gw190521}, to substantiate our results that are based on the PN approach. 

\par
It will also be interesting to explore the event rates for hyperbolic encounters by 
employing various astrophysical models, as  described in Table~1 of~\cite{mukherjee2021gravitational},
while invoking our PN-accurate orbital description.
We suspect that these rates may not be very sensitive to our post-Newtonian description, mainly because the probability of a detectable encounter is essentially proportional to the difference in the squares of the maximum and minimum values of the angle between the two masses as shown in Eq.~(18) of ~\cite{mukherjee2021gravitational} and our post-Newtonian description typically does not change these values substantially. In the most favorable scenarios, it is possible that  PN changes to the event rates are compared to model systematics as displayed in  Table 1 of ~\cite{mukherjee2021gravitational}. However, our prescription should be helpful to extract with confidence coincident detection in two observatories due to PN-induced features in our GW polarisation states.  

In the PTA-related efforts, it will be desirable to adapt the \texttt{BayesHopperBurst} package, a Bayesian search algorithm  for extracting generic GW bursts in PTA datasets, detailed in Ref.~\cite{becsy2021bayesian}, for our Burst with linear memory GW events.
Such efforts should allow us to provide astrophysical  bounds on the occurrence of hyperbolic encounters of SMBHs at cosmological distances by employing the existing and expected IPTA data releases \cite{perera2019iptadr2}.

\section*{Acknowledgements}
We thank Gihyuk Cho, Lankeswar Dey, and Philippe Jetzer  for helpful 
discussions.
SD and AG are grateful for the financial support and hospitality of the Pauli Center for Theoretical Studies and the University of Z\"{u}rich. 
SD, PR, AG, and AS acknowledge the support of the Department of Atomic Energy, Government of India, under project identification \# RTI 4002.
AS is supported by the NANOGrav NSF Physics Frontiers Center (awards \#1430284 and 2020265). 
ST is supported by the Swiss National Science Foundation (SNSF) Ambizione Grant Number: PZ00P2\_202204.
Throughout the stages of this work, MH was supported by the University of Zurich (grant number FK-21-123) and the Nederlandse Organisatie voor Wetenschappelijk Onderzoek (NWO-I). 
We acknowledge the use of the following software packages in this work: \texttt{numpy}~\cite{harris2020numpy}, 
\texttt{scipy}~\cite{virtanen2020scipy}, 
\texttt{astropy}~\cite{pricewhelan2018astropy}, \texttt{matplotlib}~\cite{hunter2007matplotlib}, \texttt{LALSuite}~\cite{lalsuite_official}, \texttt{PyCBC}~\cite{biwer2019pycbc}, and \texttt{ENTERPRISE}~\cite{ellis2019enterprise}.

\appendix

\renewcommand\thefigure{\thesection.\arabic{figure}}  
\begin{widetext}
\section{ Probing Properties of LVK-relevant BWM events }
\label{AppA}
  Our detailed numerical explorations reveal that 
 the horizon distance $(D_H)$ is a monotonic function of eccentricity (e) and impact parameter (b) while it varies in a non-monotonic manner with respect to the total source mass (M).
 This inference prompted  us to explore  how horizon distances vary as a function of  total source mass while fixing both $e$ and $b$ values.
In Fig~\ref{horizon-heatmap-fixm_special}, we explore two scenarios and in the first case (Scenario I), we let 
 $r_\text{min} \gtrsim 10 \ \zeta$
and in the Scenario II, we relax the above $r_\text{min}$  restriction.
Recall that the validity of our PN approximation to describe 
the Scenario II hyperbolic events
require further investigations as detailed earlier. Further, note that 
$r_\text{min}$  restriction imposes constraints on $e_t$ and $b$ values, as displayed 
in Fig.~\ref{fig:b_et_AR}.
We, as expected, see that the aLIGO Horizon distance reach for the Scenario II hyperbolic 
events are substantially higher compared to Scenario I events.
Interestingly, the  peak of $D_H$ shifts towards higher masses for the Scenario II compared its counterpart. We now display in Fig~\ref{horizon-cons_e-all-zone}  where we repeat what is pursued in Fig.~\ref{horizon-heatmap-conse} while relaxing the 
 restriction that $r_\text{min} \gtrsim 10 \ \zeta$.
 We observe that such hyperbolic events with $b \sim 50 \, \zeta$ and $e\sim 1.1$ can be visible to 
 $\sim 500$ Mpc for aLIGO which should  motivate further explorations of hyperbolic events 
 in full General Relativity.

\begin{figure*}
\centering  
\includegraphics[width=0.85\linewidth]{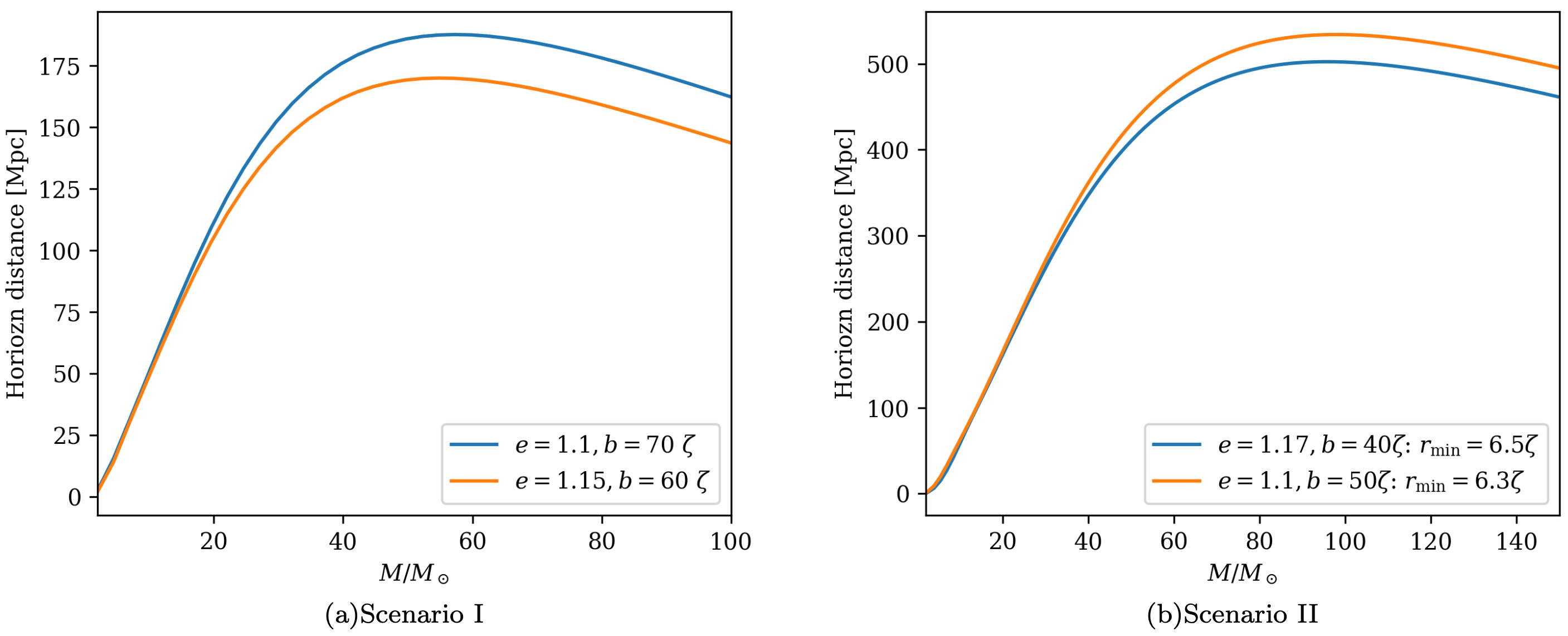}
\caption{Horizon distance plots for aLIGO with hyperbolic encounters of fixed eccentricity and impact parameter for BBH system while we keep the total mass as a variable. For the left plot, we choose two optimal configurations 
while ensuring that 
$r_\text{min} \gtrsim 10 \ \zeta$, and for the right plot, we relax this restriction but make sure that  $r_\text{min}\gtrsim 6 \ \zeta$.
Clearly, highly relativistic hyperbolic encounters can be visible up to 
$500$Mpc.
}
\label{horizon-heatmap-fixm_special}
\end{figure*}

\begin{figure*}
\centering  
\includegraphics[width=0.85\linewidth]{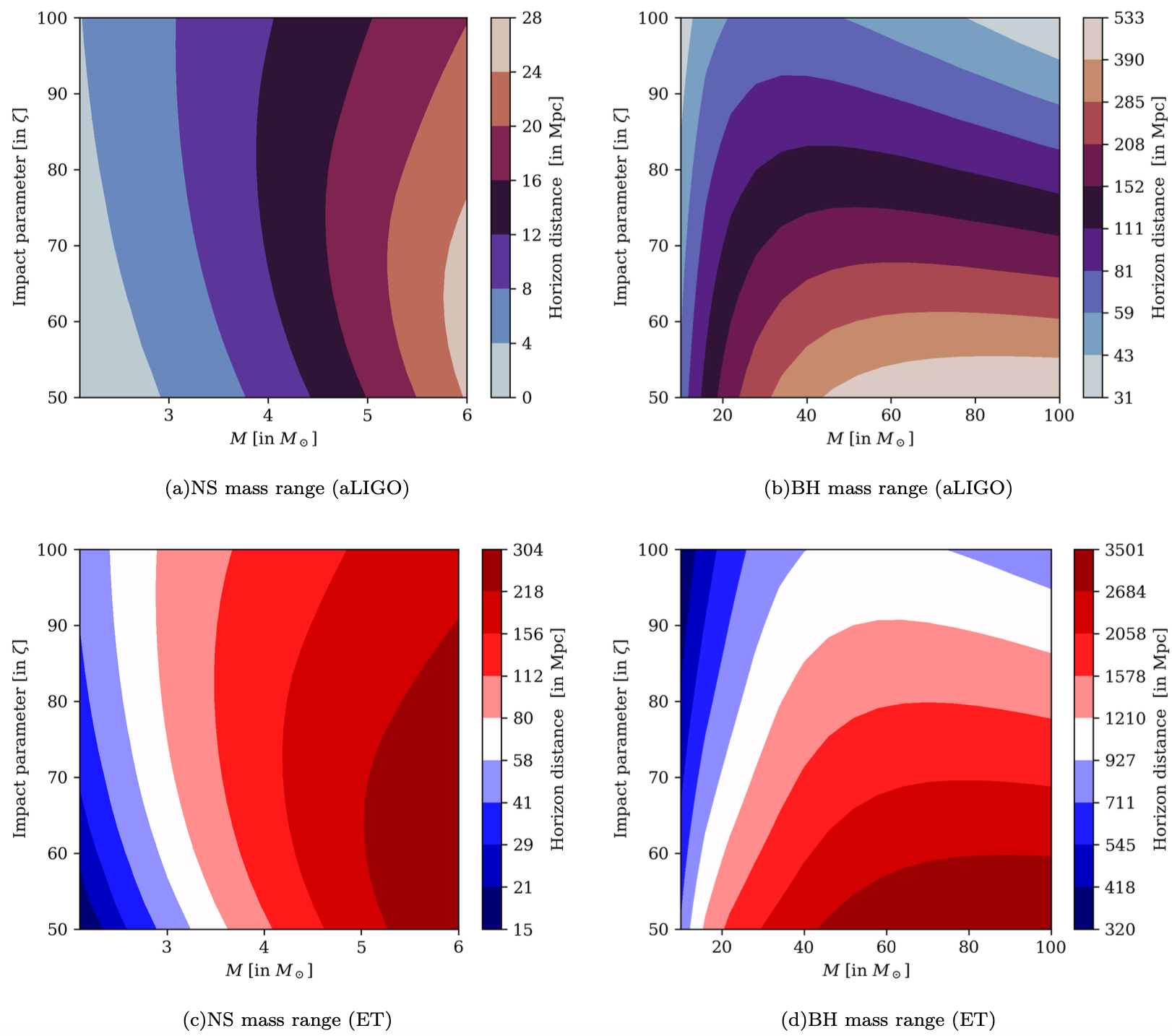}
\caption{ Contour plots for 
the horizon distances  of hyperbolic encounters where we relax the 
 restriction that $r_\text{min} \gtrsim 10 \ \zeta$.
We are focusing on equal mass systems, relevant for aLIGO and ET and let 
$e_t=1.1$. We observe that the reach of NS systems are not substantially 
changed, compared to Fig.~\ref{horizon-heatmap-conse}. However, BH binaries are visible to $\sim 500$ Mpc distances even for 
aLIGO.}
\label{horizon-cons_e-all-zone}
\end{figure*}

\section{3PN accurate \texorpdfstring{$\frac{du}{dt}$}{du/dt} in Modified Harmonic  gauge}
\label{dudt3pn}
Recall that for our \texttt{HyperbolicTD} approximant, we require to solve differential equations for $n,e_t$ and 
$u$. In what follows,
we provide explicit PN contributions to the 3PN-accuarte  $du/dt$ expression and we write 
\begin{align}
\frac{du}{dt} &=\frac{x^{3/2} \, c^3}{G \, M \, \beta} \bigg{\{} A_\text{Q}+A_\text{2PN}+A_\text{2.5PN}+A_\text{3PN}\bigg{\}}
\end{align}
where $\beta=(e_t \, \cosh u-1)$, and various PN contributions are 
\begin{subequations}
\begin{align}
A_\text{Q}&=1 \\[2ex]
A_\text{2PN}&=-\frac{x^2}{8 \, \beta^3} \bigg{[} (60-24 \, \eta)\, \beta+(15-\eta) \, \eta \, e_t \, (e_t-\cosh u) \bigg{]} \\[2ex]
A_\text{2.5PN}&=\frac{8 (1-e_t^2) \, x^{5/2}}{15 \, \beta^6} \eta \, \sinh u  \big{(}-26 e_t+24 \cosh u-9 \, e_t^2 \cosh u+8 \, e_t \cosh^2 u+3 \, 
		e_t^2 \cosh^3 u\big{)} \\[2ex]
	A_\text{3PN}&=\frac{1}{6720 (e_t^2-1)^{3/2} \beta ^3}(1-e_t) \, x^3  \cosh^2 \left(\frac{u}{2}\right) \bigg{\{}35 \, \beta  \, \gamma \, 
		\bigg{(}8640+\bigg{(}-13184+123 \pi ^2\bigg{)} \eta +960 \, \eta ^2 \notag \\[2ex] 
		&+96 e_t^2 (30-29 \eta +11 \eta ^2)\bigg{)} \sech^2 \frac{u}{2}+e_t
		\gamma  \bigg{(}67200-3 (-47956+105 e_t^2+1435 \pi ^2) \eta  \notag \\[2ex] 
		&-105 (592+135 e_t^2)\eta ^2+35 (-8+65 e_t^2)
		\, \eta ^3\bigg{)} (e_t-\cosh u) \sech^2 \frac{u}{2}-840 e_t \gamma  (-4+\eta ) \cosh u \notag \\[2ex] 
		& \times  \bigg{(}-60+3 (8+5
		e_t^2) \eta -e_t^2 \eta ^2+e_t (60-39 \eta +\eta ^2) \cosh u\bigg{)} \sech^2 \frac{u}{2}\notag \\[2ex] 
		&-\frac{420 \, 
			e_t^2 \,  \gamma  \, \eta  (116-49 \eta +3 \eta ^2) (-3 e_t^2+4 e_t \cosh u+(e_t^2-2) 
			\cosh 2u) \sech^2 \frac{u}{2}}{\beta }\notag \\[2ex] 
		&+\frac{1}{4 \beta ^2}105 e_t^3 \, \gamma  \, \eta  \, (23-73 \eta +13 \eta ^2)\bigg{(}10 \, 
		e_t^3-15 e_t^2 \cosh u-6 e_t (e_t^2-2) \cosh 2u-4 \cosh (3u)\notag \\[2ex] 
		&+3 e_t^2 \cosh 3u \bigg{)}  \times \sech^2 \frac{u}{2}-\frac{3360 \, e_t^2 \, (1+e_t) \sqrt{e_t^2-1} (-15+\eta ) (-4+\eta ) \eta  \sinh^2 \frac{u}{2}}{\beta
		}\notag \\[2ex] 
	&-3360 \, e_t^2 \, \gamma  \, (4-\eta ) \bigg{(}60-24 \eta -\frac{e_t (-15+\eta ) \eta  (e_t-\cosh u)}{\beta }\bigg{)} \sinh^2 \frac{u}{2}\bigg{\}},
	\end{align}
\end{subequations}
 and  we let $\gamma=\sqrt{\frac{e_t+1}{e_t-1}}$.
 
\end{widetext}

\bibliographystyle{apsrev4-1}
\bibliography{mybib}

\end{document}